\documentclass[longauth]{aa} 
\usepackage{graphicx}
\usepackage{txfonts}
\usepackage{color}
\usepackage[breaklinks=true,colorlinks=true,citecolor=blue,linkcolor=blue]{hyperref}

\newcommand{\thisgrb}{GRB\,130606A}

\begin{document} 

   \title{VLT/X-shooter spectroscopy of the afterglow of the {\it Swift} GRB\,130606A\thanks{Based on observations carried out under prog. ID 091.C-0934(C)
with the X-shooter spectrograph installed at the Cassegrain focus of the Very Large Telescope (VLT), Unit 2 -- Kueyen, operated by the European Southern
Observatory (ESO) on Cerro Paranal, Chile. 
Partly based on observations made with the Nordic Optical Telescope, operated on the island of La Palma jointly by Denmark, Finland, Iceland, Norway, and Sweden, in the Spanish Observatorio del Roque de los Muchachos of the Instituto de Astrof\'isica de Canarias.
Partly based on observations made with the Italian Telescopio Nazionale Galileo (TNG) operated on the island of La Palma by the Fundaci\'on Galileo Galilei of the INAF (Istituto Nazionale di Astrofisica) at the Spanish Observatorio del Roque de los Muchachos of the Instituto de Astrof\'isica de Canarias, under programme A26TAC\_63.}}

  \subtitle{Chemical abundances and reionisation at $z\sim6$}

   \author{O.~E.~Hartoog\inst{1}
          \and
          D.~Malesani\inst{2}
          \and
          J.~P.~U.~Fynbo\inst{2}
          \and 
          T.~Goto\inst{2,3}
          \and 
          T.~Kr{\"u}hler\inst{2,4}
          \and
          P.~M.~Vreeswijk\inst{5}
          \and
          A.~De Cia\inst{5}
          \and 
          D.~Xu\inst{2}          
          \and
          P.~M{\o}ller\inst{6}
          \and
          S.~Covino\inst{7}
          \and
          V.~D'Elia\inst{8,9}
          \and
          H.~Flores\inst{10}
          \and
          P.~Goldoni\inst{11}
          \and
          J.~Hjorth\inst{2}
          \and
          P.~Jakobsson\inst{12}
          \and
          J.-K.~Krogager\inst{2,4}
          \and
          L.~Kaper\inst{1}
          \and
          C.~Ledoux\inst{4}
          \and
          A.~J.~Levan\inst{13}
          \and
          B.~Milvang-Jensen\inst{2}
          \and
          J.~Sollerman\inst{14}
          \and
          M.~Sparre\inst{2}
          \and
          G.~Tagliaferri\inst{7}
          \and
          N.~R.~Tanvir\inst{15}
          \and
          A.~de~Ugarte~Postigo\inst{2,16}
          \and
          S.~D.~Vergani\inst{10}
          \and
          K.~Wiersema\inst{15}
          \and
          J.~Datson\inst{17}
          \and
          R.~Salinas\inst{18,19}
          \and
          K.~Mikkelsen\inst{20}
          \and
          N.~Aghanim\inst{21}
          }

   \institute{Anton Pannekoek Institute for Astronomy, University of Amsterdam, Science Park 904, PO Box 94249, 1090 GE Amsterdam, The Netherlands,
              \email{O.E.Hartoog@uva.nl}
         \and
         Dark Cosmology Centre, Niels Bohr Institute, Copenhagen University, Juliane Maries Vej 30, 2100 C{\o}penhagen O, Denmark
         \and
         Institute of Astronomy and Department of Physics, National Tsing Hua University, No. 101, Section 2, Kuang-Fu Road, Hsinchu 30013, Taiwan, R.O.C
        \and
        European Southern Observatory, Alonso de C\'{o}rdova 3107, Vitacura, Casilla 19001, Santiago 19, Chile
        \and
        Department of Particle Physics and Astrophysics, Faculty of Physics, Weizmann Institute of Science, Rehovot 76100, Israel
          \and
         European Southern Observatory, Karl-Schwarzschildstrasse 2, 85748 Garching bei M\"unchen, Germany
         \and
         INAF, Osservatorio Astronomico di Brera, Via E. Bianchi 46, I-23807 Merate, Italy
         \and
         INAF-Osservatorio Astronomico di Roma, Via Frascati 33, I-00040 Monteporzio Catone, Italy
        \and
         ASI-Science Data Center, Via Galileo Galilei, I-00044 Frascati, Italy   
         \and
         Laboratoire GEPI, Observatoire de Paris, CNRS-UMR8111, Univ. Paris-Diderot 5 place Jules Janssen, 92195 Meudon France
         \and
         APC, Astroparticule et Cosmologie, Universite Paris Diderot, CNRS/IN2P3, CEA/Irfu, Observatoire de Paris, Sorbonne Paris Cit\'e, 10, Rue Alice Domon et L\'eonie Duquet, 75205, Paris Cedex 13, France  
         \and
         Centre for Astrophysics and Cosmology, Science Institute, University of Iceland, Dunhagi 5, IS-107 Reykjavk, Iceland
         \and
         Department of Physics, University of Warwick, Coventry CV4 7AL, United Kingdom
         \and
         The Oskar Klein Centre, Department of Astronomy, AlbaNova, SE-106 91, Stockholm, Sweden   
         \and
         University of Leicester, Department of Physics and Astronomy, University Road, Leicester LE1 7RH, United Kingdom
         \and
         Instituto de Astrof\' isica de Andaluc\' ia (IAA-CSIC), Glorieta de la Astronom\' ia s/n, E-18008, Granada, Spain
         \and         
         Tuorla Observatory, Department of Physics and Astronomy, University of Turku, FI-20014 Turku, Finland
         \and
         Finnish Centre for Astronomy with ESO, University of Turku, V{\"a}is{\"a}l{\"a}ntie 20, 21500, Piikki{\"o}, Finland
         \and
          Department of Physics and Astronomy, Michigan State University, East Lansing, MI 48824, USA
          \and
          Institute of Theoretical Astrophysics, University of Oslo, P.O. Box 1029, Blindern, NO-0315 Oslo, Norway
          \and
          Institut d'Astrophysique Spatiale, CNRS (UMR8617) Universit{\'e} Paris-Sud 11, B{\^a}timent 121, Orsay, France
             }

   \date{Received 16-09-2014; accepted 15-06-2015}

  \abstract   
   {The reionisation of the Universe is a process that is thought to have ended around $z\sim6$, as inferred from spectroscopy of distant bright background sources, such as quasars (QSO) and gamma-ray burst (GRB) afterglows. Furthermore, spectroscopy of a GRB afterglow provides insight in its host galaxy, which is often too dim and distant to study otherwise.  }
   {For the {\it Swift} \thisgrb~at $z=5.913$ we have obtained a high S/N spectrum covering the full optical and near-IR wavelength region at intermediate spectral resolution with VLT/X-shooter. We aim to measure the degree of ionisation of the intergalactic
medium (IGM) between $z=5.02-5.84$ and to study the chemical abundance pattern and dust content of its host galaxy.}
   {We estimated the UV continuum of the GRB afterglow using a power-law extrapolation, then measured the flux decrement due to absorption at Ly$\alpha,\beta,$ and $\gamma$ wavelength regions. Furthermore, we fitted the shape of the red damping wing of Ly$\alpha$. 
   The hydrogen and metal absorption lines formed in the host galaxy were fitted with Voigt profiles to obtain column densities. We investigated whether ionisation corrections needed to be applied.  }
   {Our measurements of the Ly$\alpha$-forest optical depth are consistent with previous measurements of QSOs, but have a much smaller uncertainty. The analysis of the red damping wing yields a neutral fraction $x_\ion{H}{i}<0.05$ ($3\sigma$). We obtain column density measurements of H, Al, Si, and Fe; for C, O, S and Ni we obtain limits. The ionisation due to the GRB is estimated to be negligible (corrections $<0.03$\,dex), but larger corrections may apply due to the pre-existing radiation field (up to $0.4$\,dex based on sub-DLA studies). Assuming that $\mathrm{[Si/Fe]}=+0.79\pm0.13$ is due to dust depletion, the dust-to-metal ratio is similar to the Galactic value.}
   {Our measurements confirm that the Universe is already predominantly ionised over the redshift range probed in this work, but was slightly more neutral at $z>5.6$. GRBs are useful probes of the ionisation state of the IGM in the early Universe, but because of internal scatter we need a larger statistical sample to draw robust conclusions.
The high [Si/Fe] in the host can be due to dust depletion, $\alpha$-element enhancement, or a combination of both. The very high value of $\mathrm{[Al/Fe]=2.40\pm0.78}$ might be due to a proton capture process and is probably connected to the stellar population history. We estimate the host metallicity to be $-1.7<[\mathrm{M/H}]<-0.9$ ($2\%-13\%$ of solar). 
}

   \keywords{Gamma-ray bursts: individual: GRB\,130606A, Cosmology: observations, Cosmology: dark ages, reionisation, first stars, ISM: abundances}

   \maketitle

\section{Introduction}
The potential of gamma-ray bursts (GRBs) as probes of star formation and the physics of the inter-galactic medium (IGM) back to the epoch of the first galaxies was appreciated very early in the so-called afterglow era \citep[e.g.,][]{1998MNRAS.294L..13W,2000ApJ...536....1L,2000ApJ...540..687C}. This promise has been bolstered by the rapid increase in the highest recorded spectroscopic redshift for GRBs from $z=4.5$ in 2000 to $z=8.2$ in 2009 \citep{2000A&A...364L..54A, 2009Natur.461.1254T,2009Natur.461.1258S}. There are three main objectives in using GRBs to probe star formation and the IGM at high redshifts: to measure chemical abundances in very ``primitive'' conditions \citep[e.g., dominated by Population III stars,][]{2007ApJ...663L..57P,2012ApJ...760...27W,2014ApJ...785..150S}, to measure the neutral fraction of the IGM from the shape of the red damping wing \citep[e.g.,][]{1998ApJ...501...15M,2006PASJ...58..485T, 2009ScChG..52.1428X,2009ApJ...693.1610G,2010A&A...512L...3P}, and to pinpoint locations of star formation \citep{2007ApJ...665..102B,2007ApJ...669....1R,2012ApJ...754...46T}. The last objective advances with each new detection, the full potential is still far from realised especially for the first two objectives. 

GRB afterglows can be extremely bright, but also fade very rapidly. To obtain high-quality spectra, rapid follow-up at the largest optical telescopes is required. Primarily thanks to the \emph{Swift} mission \citep{2004ApJ...611.1005G}, studying the high-redshift universe though GRB afterglows has progressed well over the last years \citep[e.g.,][]{2006Natur.440..184K,2007ApJ...663L..57P,2007ApJ...669....1R,2013MNRAS.428.3590T,2014ApJ...785..150S}. Traditionally, high-redshift galaxies are selected as Lyman-break galaxies \citep[LBGs,][]{2002ARA&A..40..579G,2003ApJ...592..728S}, Ly$\alpha$ emitters \citep{2010MNRAS.408.1628S,2012ApJ...744...83O,2014arXiv1403.5466P}, and damped Ly$\alpha$ absorbers towards quasars \citep[QSO-DLAs,][]{2005ARA&A..43..861W}. GRB host galaxies studied in absorption through afterglows (sometimes referred to as GRB-DLAs) are analysed using similar methods as those applied to QSO-DLAs, but intrinsic and observational biases are very different between these two classes of objects \citep[see e.g.][]{2007ApJ...666..267P,2008ApJ...683..321F}. In contrast to LBGs, for example, GRB host galaxies are not selected by their brightness; this provides the opportunity to explore a different range of luminosities and therefore sizes and masses of high-redshift galaxies.

We present spectroscopic observations of the afterglow of the $z=5.913$ GRB\,130606A obtained with X-shooter on the European Southern Observatory (ESO) Very Large Telescope (VLT). The objective of the paper is to address the questions of chemical abundances in the host galaxy and the ionisation state of the IGM.
Independent analyses of this event used data of significantly lower spectral resolution and with smaller wavelength coverage,
for example, \citet{2013ApJ...774...26C}, \citet{2013arXiv1312.5631C} and \citet{2014PASJ...66...63T}. Our results are discussed in the context of the results from these earlier reports. The superior resolution and wavelength range we achieve with the data presented here, with the associated ability to perform direct line fits for abundances, provide the motivation for this paper.

For the cosmological calculations we assume a $\Lambda$CDM universe with $\Omega_\Lambda = 0.6911$, $\Omega_\mathrm{m} = 0.3089$, and $H_0=67.74$~km~s$^{-1}$ ~Mpc$^{-1}$ from the 2013 \emph{Planck} data \citep{2014A&A...571A..16P}. Magnitudes are given in the AB system throughout the paper. Column densities are given as $\log(N/\mathrm{cm^{-2}})$. We use $1\sigma$ error bars unless explicitly noted otherwise.

\section{Observations}

\begin{table}
\fontsize{8}{10}\selectfont
\caption{Top: Log of the NOT and TNG observations with (1) mid-time of observations, (2) time since the burst (observer's frame), (3) exposure time, (4) seeing, (5) filter, and (6) measured magnitude. Bottom: main parameters of the VLT/X-shooter observations, frame by frame, with (1) mid-time, (2) exposure time, (3) average airmass,
and (4) average seeing measured in the VIS and NIR 2D spectra. The last line contains the average values for the VLT/X-shooter observations.}
\label{tab:log}
\center
\begin{tabular}{cccccc}
\multicolumn{6}{c}{NOT and TNG}\\
\hline
Mid-time & ${T_\mathrm{GRB}}$ & Exptime  & Seeing & Filter & Magnitude \\
  UTC      & (day)        & (s)         &   (")     &        &  (AB mag)  \\  
\hline
June 6.89179  & 0.01356 &  30    &    1.6    & NOT/$r$  & $20.51\pm0.06$\\
June 6.89434  & 0.01611 &  30    &    1.6    & NOT/$r$  & $20.79\pm0.06$\\
June 6.89727  & 0.01904 &  300  &    1.6    & NOT/$r$  & $20.90\pm0.03$\\
June 6.90137  & 0.02314 &  300  &    1.8    & NOT/$r$  & $21.19\pm0.04$\\
June 7.02835  & 0.15012 &  300  &    1.1    & NOT/$i$   & $19.83\pm0.03$\\
June 7.03241  & 0.15418 &  300  &    1.1    & NOT/$r$  & $22.93\pm0.08$\\
June 7.90616  & 1.02793 & 5$\times$300&    0.7    & NOT/$z$  & $21.02\pm0.03$\\
June 7.92510  & 1.04687 & 4$\times$300&    0.7    & NOT/$i$  & $23.37\pm0.08$\\
June 8.15646  & 1.27823 & 7$\times$180&   1.1    & TNG/$i$  & $23.99\pm0.14$\\
June 8.17616  & 1.29793 & 7$\times$180&   1.1    & TNG/$z$  & $21.50\pm0.04$\\
June 10.1230  & 3.24474 & 3$\times$600&   0.8    & TNG/$z$  & $22.26\pm0.14$\\
June 12.0047  & 5.12648 & 8$\times$300&   0.6    & NOT/$z$  & $>22.70$\\
\hline
\\
\end{tabular}
\begin{tabular}{cccccc}

\multicolumn{5}{c}{VLT/X-shooter}\\
\hline
Mid-time  & ${T_\mathrm{GRB}}$& Exptime  & Airmass & Seeing \\
  UTC      & (day) & (s)  &         & (")  \\  
\hline
June 7.17990 & 0.30167 & 600 & 1.72  & 1.2 - 0.9 \\
June 7.18809 & 0.30986 & 600 & 1.72  & 1.2 - 0.9 \\
June 7.19899 & 0.32076 & 600 & 1.73  & 1.2 - 0.9 \\
June 7.20707 & 0.32884 & 600 & 1.74  & 1.2 - 1.0\\
June 7.23147 & 0.35324 & 600 & 1.77 & 1.4 - 1.2\\
June 7.23955 & 0.36132 & 600 & 1.88  & 1.1 - 0.9 \\
\hline
\multicolumn{5}{c}{Average values} \\
\hline
June 7.20751 &0.32928 & 600 & 1.76  & 1.22 - 0.97 \\
\hline

\end{tabular}
\end{table}

GRB\,130606A was detected by {\it Swift} on June 6, 2013 at 21:04:39 UT \citep{2013GCN..14781...1U}. The burst was relatively long with a $T_{90}$ duration of $277\pm19\,\rm s$ \citep{2013GCN..14819...1B}. 

\begin{figure}
\centering
\includegraphics[width=8.2cm,clip]{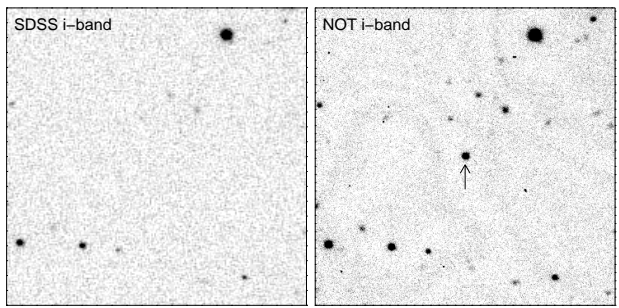}
\caption{
$80''\times80''$ field around the position of GRB\,130606A in the $i$ band. The left panel shows a pre-explosion image from the Sloan Digital Sky Survey and the right panel our $i$ band image obtained with the NOT 3.5 hr after the burst. North is up and east is to the left.
}
\label{Afterglow}
\end{figure}

We first observed the field of GRB\,130606A with the Nordic Optical Telescope (NOT) equipped with the MOsaic CAmera (MOSCA) in the $r$ band (3$\times$200\,s). Table~\ref{tab:log} lists the observation log, including those obtained with the Telescopio Nazionale Galileo (TNG). Observations started at 20:35\,UT or 0.5 hr after the GRB trigger \citep{2013GCN..14781...1U,2013GCN..14783...1X}. Consistent with the \emph{Swift}/XRT position \citep{2013GCN..14811...1O}, we detected a new source with an $r$-band magnitude of 20.8\,mag at RA (J2000) = 16$^\mathrm{h}$37$^\mathrm{m}$35.188$^\mathrm{s}$, Dec (J2000) =+29$^{\circ}$47$^{\prime}$47.03$^{\prime\prime}$ \citep{2013GCN..14783...1X}. The source coincided with the one reported by \citet{2013GCN..14782...1J}. Observations in the near-infrared revealed that the afterglow was extremely bright \citep[$K_\mathrm{s}=15.0$ about half an hour after the burst;][]{2013GCN..14794...1N,2013GCN..14784...1N}. We observed the field again 3\,hr later with NOT/MOSCA in the $r$ and $i$ bands. In Fig.~\ref{Afterglow} we show our $i$-band image compared to a pre-explosion image from the Sloan Digital Sky Survey. It soon became clear through spectroscopy and multi-band photometry at a number of telescopes that GRB\,130606A was a very distant GRB \citep{2013GCN..14796...1C,2013GCN..14798...1L,2013GCN..14807...1A,2014AJ....148....2L}.  We subsequently acquired a medium-resolution spectrum with the X-shooter spectrograph mounted at the ESO/VLT \citep{2011A&A...536A.105V}, using nodding mode with $1\times2$ binning \citep[i.e., binning in the dispersion direction,][]{2013GCN..14816...1X}. Because we used a $K$-band blocking filter to increase the signal-to-noise ratio (S/N) at shorter near-IR wavelengths, the spectral coverage extends from 3000 to about 20\,000\,\AA, corresponding to $430-2900\,\AA$ in the GRB rest frame. Observation started at 03:57:41 UT on 7 June, 2013 (see Table~\ref{tab:log}). Because the atmospheric dispersion corrector was not working at the time of the observation, we aligned the slit according to the parallactic angle and spread the observation over three observing blocks (OBs) of $2\times600\,\rm s$ each in the UVB, VIS, and NIR arm. After each OB, we reset the position angle to the new parallactic angle to minimise flux losses in the UVB and VIS arm. The mid-exposure time is 7.829\,hr post burst \citep{2013GCN..14816...1X}. The slit widths were matched to the seeing conditions, that is, we chose a 1\farcs0, 0\farcs9, and 0\farcs9 slit in the UVB, VIS, and NIR arm, respectively. For this given instrument setup, the nominal resolving power $R=\lambda/\Delta \lambda$ is 5100, 8800, and 5300 in the UVB, VIS and NIR, respectively.  
For the VIS and NIR spectra we were able to directly measure the resolving power from the width of telluric absorption lines and find it to be 8700 and 6200, respectively. The NIR resolving power is higher than the nominal one because during most of the observations the seeing in this wavelength range was smaller than the slit width. The UVB spectrum shows no afterglow signal, as expected, given that the coverage of this arm falls below the Lyman break at the redshift of the GRB, and is not discussed further in the remainder of this paper.

VLT/X-shooter data were reduced with the X-shooter pipeline version {\tt2.2.0}\footnote{\tt http://www.eso.org/sci/software/pipelines/} \citep{2011AN....332..227G}. The wavelength binning was chosen to be 0.2 \AA/px in the VIS, and 0.6 \AA/px in the NIR.
All spectra were flux calibrated with the spectrophotometric standard star LTT3218. We transformed the wavelength solution to vacuum and to the heliocentric frame.
We corrected the VIS and NIR spectra for telluric absorption using the spectra of the telluric standard star Hip095400 observed just after the afterglow with the same slit width and at a similar airmass. The telluric corrections built with those spectra were applied with the {\sc Spextool} software \citep{vacca03}. This corrected version was only used for line measurements in contaminated regions, because the uncorrected results showed a slightly higher S/N in unaffected regions.

\section{Results}

\subsection{Astrometry}

Based on the $i$-band image shown in Fig.~\ref{Afterglow}, we determine the position of the afterglow of RA(J2000) = 16:37:35.143, Dec(J2000) = +29:47:46.62 calibrated to the USNO-A2.0 catalogue \citep{monet1998}. 
The estimated error on the absolute position is 0.3" \citep{1999AJ....118.1882D}. The position relative to stars in the field is much more precise, with an uncertainty of about 0.05".

\begin{figure}
\centering
\includegraphics[angle=90,width=9cm,clip]{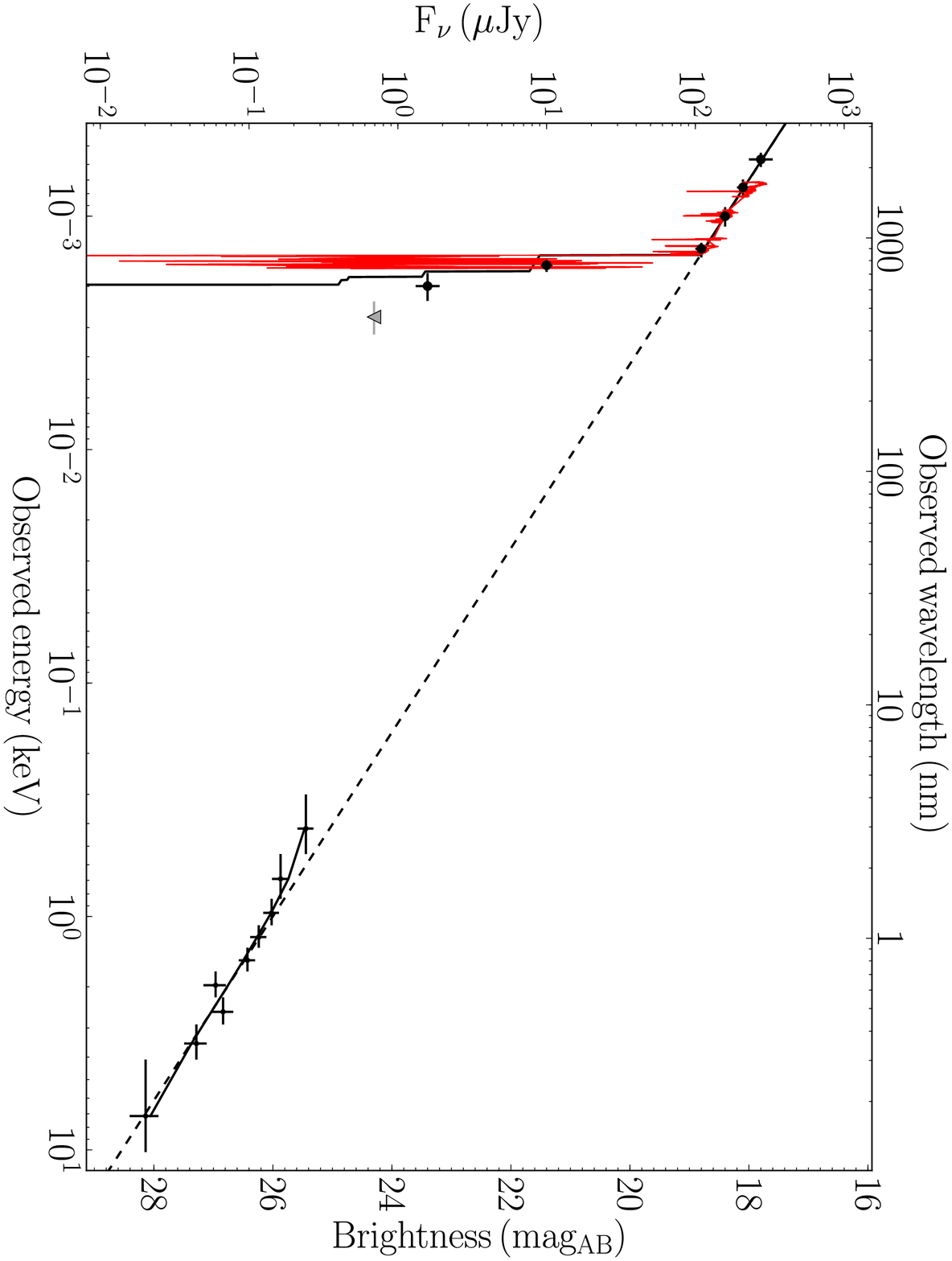}
\caption{
Afterglow spectral energy distribution. The red spectrum is the X-shooter spectrum, the filled circles on the left are the GROND photometric points \citep{2013GCN..14807...1A}, and the data points on the right are the (binned) \emph{Swift}/XRT points.  The solid line is the synchrotron spectrum with interstellar absorption, while the dashed line shows the unabsorbed synchrotron spectrum with best fit $\beta=1.02$.
\label{Afterglow:SED}}
\end{figure}

\subsection{Fitting the spectral energy distribution}
\label{sec:sed}
We constructed a broad-band spectral energy distribution (SED) for the afterglow of \thisgrb~spanning the wavelength range from the NIR $K$ band to the X-ray energy range. We followed the methodology outlined in \citet{2011A&A...534A.108K} and used the X-shooter spectroscopy, photometric data from the Gamma-Ray Burst Optical/Near-Infrared Detector (GROND) as given in \citet{2013GCN..14807...1A} and public X-ray data from the \emph{Swift}/X-Ray Telescope (XRT) repository \citep{2007A&A...469..379E, 2009MNRAS.397.1177E}. We corrected for the Galactic foreground extinction of $A_{V,\textrm{Gal}}=0.08$ \citep{1998ApJ...500..525S,2011ApJ...737..103S}.

After scaling the NIR spectroscopy to the photometry (data bluewards of Ly$\alpha$ were not fitted) to account for slit losses and different observing times, the complete data set is well described with a single power-law continuum $F_{\nu} \propto \nu^{-\beta}$ with index $\beta=1.02\pm0.03$, without evidence for reddening towards the GRB in addition to the Galactic component (see Fig.~\ref{Afterglow:SED}). 

We set a 3$\sigma$ upper limit of $A_V < 0.2\,\rm{mag}$ at $\sim z_{\rm{GRB}}$ assuming local extinction laws from the SMC, LMC, or MW \citep{1992ApJ...395..130P}. The intrinsic X-ray absorption is also consistent with zero; given the GRB's high redshift, we obtain a $3\sigma$ upper limit of $N_{\rm{H}}^{X} < 3 \times 10^{22}\,\rm{cm}^{-2}$. 

\subsection{Analysis of the red damping wing}
\label{sec:reddamping}
\begin{figure}
\centering
\includegraphics[width=8.8cm,clip]{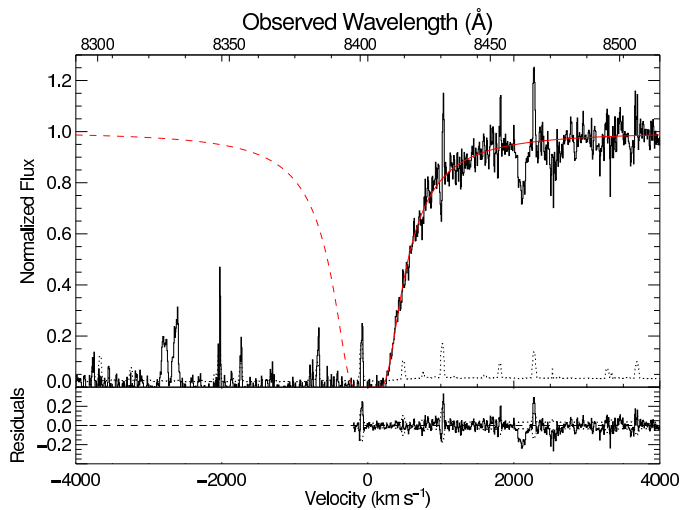}
\caption{Top panel: The spectral region around the Ly$\alpha$ feature with our best fit to the red wing overplotted in red. The model consists of a fully ionised intergalactic medium and a column density in the GRB host of $\log N_\ion{H}{i}=19.91\pm0.02$. The $1\sigma$ noise is plotted as a dotted line. The bottom panel shows the residuals from the fit, with the $\pm1\sigma$ region marked with a dotted line. The skyline residual in the bottom of the Ly$\alpha$ line should not be interpreted as Ly$\alpha$ emission.}
\label{lyafig}
\end{figure}

\begin{figure}
\centering
\includegraphics[width=8.8cm,clip]{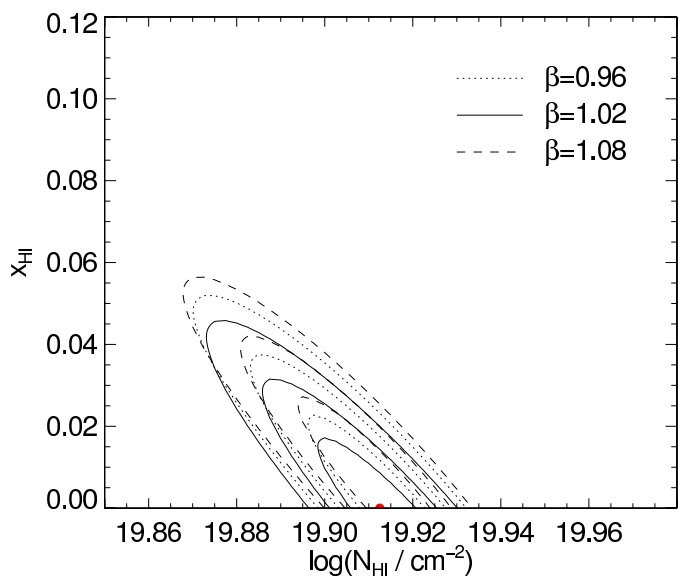}
\caption{
1, 2 and 3$\sigma$ contours for a joint fit of the neutral hydrogen column density in the host galaxy, $\log{N_{\ion{H}{i}}}$, and the neutral fraction of the intergalactic medium, $x_{\ion{H}{i}}$. The fit has been performed for three different normalising power-law SEDs: $\beta = 0.96$, $\beta=1.02$ (best fit), and $\beta=1.08$. The red dot indicates the peak likelihood.}
\label{Contour}
\end{figure}

We used the SED obtained in Sect.~\ref{sec:sed} to normalise the spectrum. The normalisation is generally much more secure for GRB afterglows, which are intrinsically simple power-laws (pure synchrotron emission), compared to the more complex spectra of QSOs. To constrain the ionisation state of the IGM, we followed \citet{1998ApJ...501...15M} and \citet{2006PASJ...58..485T} and jointly fitted the hydrogen column density in the GRB host galaxy, $\log{N_{\ion{H}{i}}}$, and the neutral fraction of the IGM, $x_{\ion{H}{i}}$. In this analysis the redshift was kept fixed to that of the strongest component of the metal lines ($z=5.91285$, see Sect.~\ref{sec:metallines} and Table~\ref{table:voigt}), and we assumed a constant neutral fraction $x_{\ion{H}{i}}$ between $z=5.8$ (the results are not very sensitive to this value) and $z=5.91285$. The $b$ parameter was also kept fixed in this fit. Each model specified by $\log{N_{\ion{H}{i}}}$ and $x_{\ion{H}{i}}$ was normalised to the observed spectrum at 8730\,\AA. The $\chi^2$ sum was calculated over the region 0 to 2000\,km\,s$^{-1}$ with respect to Ly$\alpha$.
We performed the fit for three different slopes of the underlying afterglow continuum, $\beta = 0.96, 1.02, 1.08$ corresponding to the $\pm3\sigma$ allowed region for the spectral slope (Sect.~\ref{sec:sed}). The minimum $\chi^2$ is reached for a fit with $x_{\ion{H}{i}} = 0$ ($x_{\ion{H}{i}} < 0.05$ at 3$\sigma$ significance) and $\log N_{\ion{H}{i}}=19.91\pm0.02$. The best fit to the Ly$\alpha$ red wing is shown in Fig.~\ref{lyafig} and the 1, 2, and 3$\sigma$ confidence regions in Fig.~\ref{Contour}.

\subsection{Metal absorption lines}
\label{sec:metallines}
In Figs.~\ref{fig:spec_part1} and \ref{fig:spec_part2} we show the afterglow spectrum redwards of the Ly$\alpha$ absorption line at $8400\,\AA$ up to $18\,000\,\AA$, which marks the end of the $H$ band. The spectrum has a relatively high S/N ($\sim20$ per 0.2\,\AA~pixel in VIS redwards of Ly$\alpha$; $\sim10$ per 0.6\,\AA~pixel in NIR), and many metal absorption lines are detected. Most of these are from the host galaxy, but there is a number of intervening absorbers, see Sect.~\ref{sec:int}.
 
To determine the host-galaxy metal abundances, we proceeded with Voigt-profile fits to the metal lines around $z=5.913$, with VPFIT version 10.0\footnote{\url{http://www.ast.cam.ac.uk/~rfc/vpfit.html}}. A Voigt-profile fit to the red wing of the Ly$\alpha$ absorption lines results in $\log{N_{\ion{H}{i}}} = 19.925\pm0.019$; the blue wing is not visible due to ionisation in the IGM (see Sect.~\ref{sec:IGM}). This is consistent with the value from the combined fit for $\log{N_{\ion{H}{i}}}$ and  $x_{\ion{H}{i}}$ (Sect.~\ref{sec:reddamping}); for the remainder of this analysis we adopt $\log{N_{\ion{H}{i}}} =19.91\pm0.02$. Following the $\log N_{\ion{H}{i}}$ distinctions for QSO absorbers, this system is formally a sub-DLA, but only a factor of two below the DLA threshold \citep[e.g.,][]{2005ARA&A..43..861W}.

\begin{figure*}
\centering
\includegraphics[width=16cm]{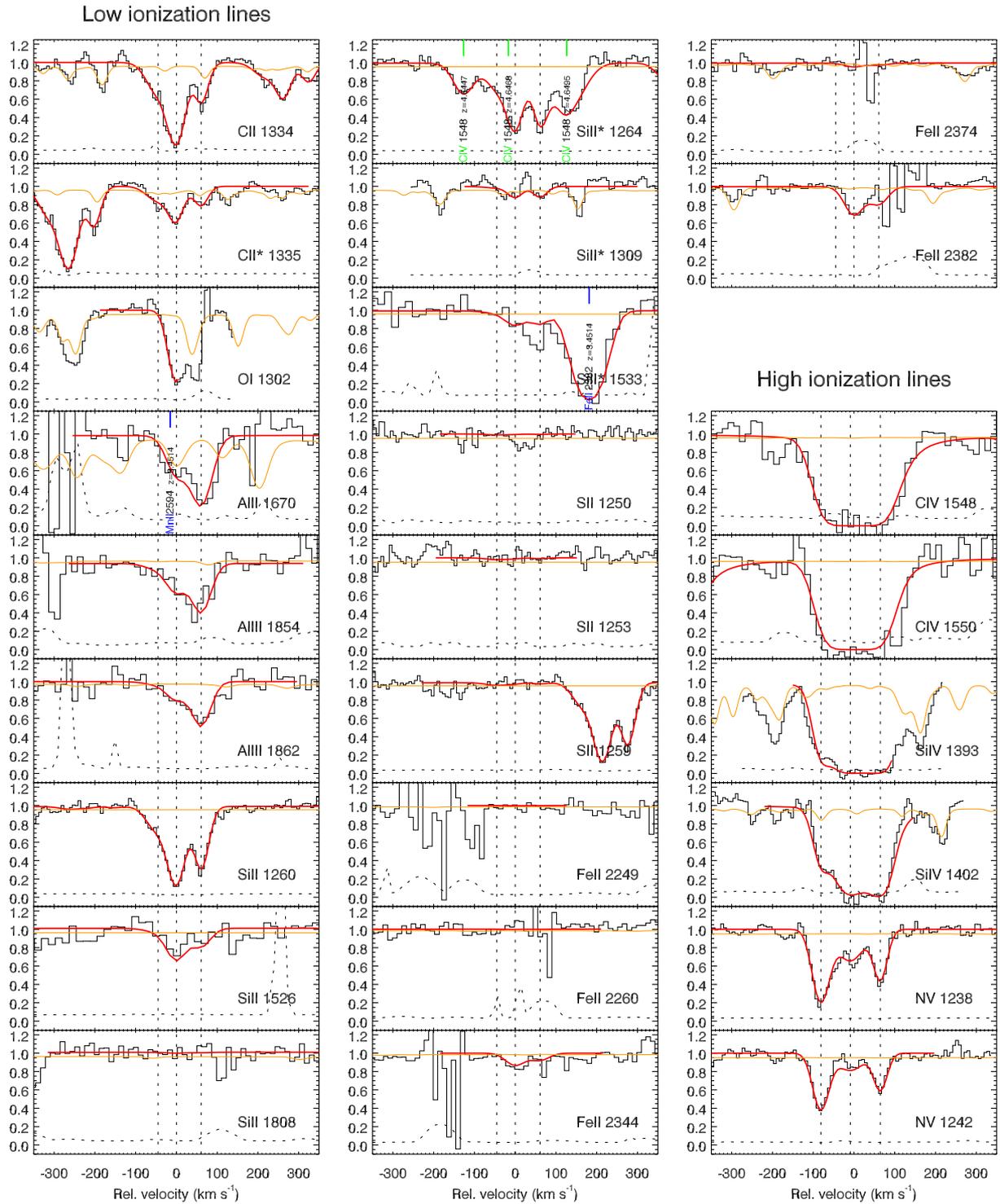}
\caption{Voigt-profile fits to the host-galaxy metal lines. We label the intervening metal lines (see Sect.~\ref{sec:int}) that contaminate the host galaxy lines; their contribution to the observed profile is taken into account. In orange we show an atmospheric transmission spectrum matching the observing conditions and spectral resolution of our observations, generated with ESO's SkyCalc tool. Residuals such as those close to the \ion{Fe}{ii} lines are due to subtracted telluric emission lines.  
\label{fig:linefits}}
\end{figure*}

We detected metal absorption lines of both low and high ionisation. We assumed that the redshift $z$ and Doppler parameter $b$ of a specific velocity component is the same for all metal lines of neutral, singly, and doubly ionised species. This assumption is commonly made for absorption spectra at intermediate spectral resolution. The $z$ and $b$ of components in the \ion{N}{v} lines were not constrained to be the same as those of the low-ionisation species because these species are expected to reside in different locations (see Sect.~\ref{sec:relstrengths}). The resulting Voigt-profile fits are shown in Fig.~\ref{fig:linefits}. \ion{C}{iv} and \ion{Si}{iv} are not included in the overall fit with free $z$ and $b$ because they are strongly saturated, but we assumed the same velocity structure as \ion{N}{v} and left the column densities free to fit. They are shown for completeness. Table~\ref{table:voigt} shows the redshifts and $b$ parameters of the Ly$\alpha$ line and the components of the metal lines that follow from the fits. Because the values of $b$ indicate that the velocity broadening is mostly due to turbulent motion of the gas and not to temperature, we did not weight the $b$ values with the ion mass. We also included undetected lines, as long as they were located in regions without strong telluric contamination, in fitting the ensemble of lines to constrain the $b$ values and reduce the effect of (hidden) saturation. We took into account contributions from intervening absorbers; for details on these, see Sect.~\ref{sec:int}.

Table~\ref{table:clumns} shows the resulting column densities for several ion species and states, per component, total, and in the metallicity notation $[\mathrm{X/H}] \equiv \log( N ( \mathrm{X})/N(\mathrm{H}))_{\mathrm{GRB}}-\log(n(\mathrm{X})/n(\mathrm{H}))_{\odot}$ based on the total column density and using reference solar abundances $n$ from \citet{2009ARA&A..47..481A} following the recommendations by \citet{2009LanB...4B...44L}. These numbers are not corrected for ionisation or dust-depletion effects; for these, see Sects.~\ref{sec:ioncor} and \ref{sec:dustmetal}, respectively.

\subsubsection{Kinematic structure of absorption lines}
\label{sec:relstrengths}
Table~\ref{table:voigt} lists the redshifts and $b$ parameters of the three components of low- (2, 4, 5) and high-ionisation lines (1, 3, 6) and their relative velocity. In the \ion{C}{ii}, \ion{Si}{ii,} and \ion{Fe}{ii} lines component 4 is stronger than 5, while this is opposite in \ion{Al}{ii} and \ion{Al}{iii}, which show structures that are very similar to one another. \ion{N}{v} shows a very different structure, which is broader, and where the absorption is strongest at the highest relative velocities. The \ion{N}{v} lines are not significantly affected by lines from intervening absorbers (see Sect.~\ref{sec:int}). High-ionisation lines such as \ion{N}{v} are common in GRB afterglow spectra \citep[see, e.g.,][]{2008A&A...491..189F}. \citet{2008ApJ...685..344P} found that six out of seven GRB afterglow spectra show \ion{N}{v} absorption, and the majority has $N(\ion{N}{v})\geq 10^{14}\mathrm{cm}^{-2}$; for GRB\,130606A we find $\log(N(\ion{N}{v})/\mathrm{cm}^{-2})=14.59\pm0.03$. However, these features are usually kinematically "cold" in GRB afterglow spectra: the lines are narrow and have velocity offsets $\delta v \lesssim 20 \mathrm{\,km\,s}^{-1}$ with respect to the location of the neutral gas. Our resolution is not high enough to make strong statements about the width of the lines, but we do see a much larger offset:  $\delta v \gtrsim 60 \mathrm{\,km\,s}^{-1}$ for the bulk of the  \ion{N}{v} absorption. In QSO-DLAs the detection rate of \ion{N}{v} is much lower, and the lines are weaker \citep{2007A&A...465..171F,2007ApJS..171...29P}.

\begin{figure}
\centering
\includegraphics[width=8.4cm,clip]{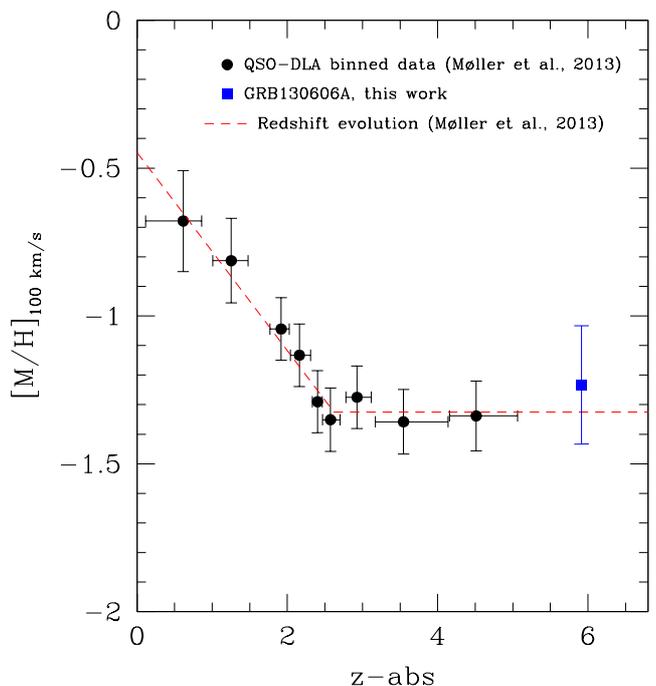}
\caption{Results from \citet{2013MNRAS.430.2680M} with the high-redshift measurement of GRB\,130606A included. [M/H]$_{100\,\mathrm{km/s}}$ is the metallicity expected for DLAs with $\Delta V = 100 \mathrm{\,km\,s}^{-1}$, based on the overall mass-metallicity relation. The data of \citet{2013MNRAS.430.2680M} (black filled circles) favour a break in the evolution of the mass-metallicity relation at $z\sim2.6$ (dashed red line), which is also supported by our measurement of GRB\,130606A (blue square). We assumed for this point $\Delta V=90 \mathrm{\,km\,s}^{-1}$ and $[\mathrm{M/H}]=-1.3\pm0.2$; see text.
\label{fig:deltav}}
\end{figure}

For QSO-DLAs it is known that the velocity width of optically thin lines, $\Delta v_{90}$, is sensitive to the stellar mass of the host galaxy \citep{2014MNRAS.445..225C};
together with the metallicity of the gas, it can therefore be used to explore the evolution of the mass-metallicity (MZ) relation of galaxies from high redshifts to the local Universe \citep[e.g.,][]{2006A&A...457...71L}. In a recent study, \citet{2013MNRAS.430.2680M} found that the redshift evolution of the MZ relation for QSO-DLAs is flat in the early Universe, but that it features a break at a redshift $z\sim2.6$ after which it evolves steeply (see Fig.~\ref{fig:deltav}). \citet{2013ApJ...769...54N} reported a
slightly flatter slope, but found no evidence for a break. 

While QSO-DLAs sample complete sightlines through galaxies, GRB-DLAs are found inside galaxies, and therefore they sample (on average) only half sightlines. \citet{2015MNRAS.446..990A} recently studied how this might alter the relations known from intervening DLA studies. Based on all the possible tests they could perform,
they concluded that all evidence indicates that GRB-DLAs follow the same relations as QSO-DLAs. They also found from their sample
that the evolution including a break \citep{2013MNRAS.430.2680M} presented a somewhat better fit (probability 100:6) than the flatter evolution with no break \citep{2013ApJ...769...54N}.

Using the definition by \citet{1997ApJ...487...73P}, $\Delta v_{90}$ is the width in velocity space containing $90\%$ of the total optical depth of a line. For GRB\,130606A we measure $\Delta v_{90}=120\pm1 \mathrm{\,km\,s}^{-1}$ from the \ion{Si}{ii}\,$\lambda1526$ transition, which is the best line available for this analysis according to the criteria set by \citet{2006A&A...457...71L}. For a wider selection of lines (loosening these criteria) we find on average $\Delta v_{90} = 140 \pm 25 \mathrm{\,km\,s}^{-1}$, which agrees with the value of \ion{Si}{ii}\,$\lambda1526$. We applied the appropriate resolution correction using the method described by \citet{2015MNRAS.446..990A} and obtained an intrinsic $\Delta v_{90}=90\mathrm{\,km\,s}^{-1}$. \thisgrb~is the highest redshift object for which $\Delta v_{90}$ is determined, and it is therefore
well suited to determine whether there is a break in the evolution of the MZ relation or not. We have included \thisgrb~in Fig.~\ref{fig:deltav}.
Keeping in mind that this is only a single point of a relation that has a significant internal scatter (0.38 dex in [M/H]), the point is clearly consistent with the evolution \emph{with} a break, as reported in
\citet{2013MNRAS.430.2680M}. The metallicity adopted here is $[\mathrm{M/H}]=-1.3\pm0.2$\footnote{This value results from our discussion in Section~\ref{sec:disc:metabun} where we take into account possible effects of ionisation and dust depletion. We assume that the estimated range $-1.7<\mathrm{[M/H]}<-0.9$ represents a $2\sigma$ confidence interval and that the metallicity is in the centre.}. The relation reported by \citet{2013ApJ...769...54N} (less steep and with no break) predicts a much lower metallicity for this $\Delta v_{90}$: $\mathrm{[M/H]}=-2.25\pm0.75$. This value marginally agrees with our metallicity measurement, but only as a result of the large scatter. We repeated the analysis of \citet{2015MNRAS.446..990A} including our
new data point at $z=5.9$ and found that considering only GRB-DLAs, the relative probability is 100:3 in favour of the redshift evolution with a break at $z=2.6$.

\subsubsection{Fine-structure and metastable lines}
Absorption lines from \ion{Si}{ii} $^2P^\circ_{3/2}$ (\ion{Si}{ii}$^{*}$) are regularly detected in high-$z$ GRB spectra \citep[e.g.,][]{2004A&A...419..927V,2006Natur.440..184K,2014ApJ...785..150S}
and were also detected in \thisgrb. They indicate that a considerable fraction of these ions are in an excited state (32\% of total \ion{Si}{ii}; see Table~\ref{table:clumns}). However, fine-structure and metastable lines from \ion{Fe}{ii} and \ion{Ni}{ii}, which are also usually present in high S/N GRB afterglow spectra \citep[see, e.g.,][]{vreeswijk2007,hartoog2013}, were not convincingly detected here. Assuming that these lines have the same velocity structure (i.e. keeping $z$ and $b$ fixed) as the resonance lines, we estimated upper limits on the gas-phase column densities of \ion{Fe}{II}\,$^6D_{7/2}$ (\ion{Fe}{ii}$^*$) and \ion{Ni}{II}\,$^4D_{9/2}$ (\ion{Ni}{ii}$^*$), which are listed in Table~\ref{table:clumns}.  These lines are probably absent because the host of \thisgrb~is a relatively weak absorber for a GRB host (not even a DLA), and the populations of the ground state levels of these excited ions are already very small due to the combined effect of a low gas column and dust depletion (Sect.~\ref{sec:dustmetal}).

\subsubsection{Ionisation correction}
\label{sec:ioncor}
Because the neutral hydrogen column density of the GRB\,130606A absorber is relatively low (sub-DLA), the hydrogen and metals are less shielded from ionising photons than in a DLA. As a result, the assumption that the overwhelming majority of an element is in a single neutral or lowly ionised state (e.g., $N_\ion{Si}{II}/N_\ion{H}{i} =
N_\mathrm{Si}/N_\mathrm{H}$), which is valid for DLAs \citep{2001ApJ...557.1007V,2002PASP..114..933P}, may not hold for this system. Moreover, the interstellar UV radiation field in a GRB host galaxy does not have to be similar to that in a QSO-sub-DLA, since the GRB host is known to be actively forming massive stars, while this is not necessarily the case in a random foreground line-of-sight object causing the QSO-sub-DLA. \citet{2009ApJ...691..152C} estimated the UV radiation field in a sample of 15 GRB host galaxies to be tens to hundreds of times that of the Galactic value, while the radiation field in QSO-DLAs, whose counterparts are notoriously difficult to detect, appears to be much more moderate \citep[e.g.,][]{2003ApJ...593..215W}.  The ionisation induced by the UV/X-ray afterglow of the GRB needs to be
considered in addition to this pre-burst ionisation correction.  Both the pre- and post-burst ionisation effects are investigated in this section.

To estimate the pre-burst ionisation correction, we compared the situation with self-consistent sub-DLA ionisation studies from the literature. \citet{2003MNRAS.345..447D} reported on a detailed study with a sample of QSO-sub-DLAs and used the logarithmic ratio $[\ion{Al}{ii}/\ion{Al}{iii}]$ or $[\ion{Fe}{ii}/\ion{Fe}{iii}]$ as an
indicator for the degree of ionisation to calibrate the ionisation parameter $U$ in their photoionisation model. The conclusion of this work is that sub-DLAs need small corrections $\lesssim0.2$\,dex for all measured elements except Al and Zn, which are more strongly influenced. For GRB\,130606A, we measure $[\ion{Al}{ii}/\ion{Al}{iii}]=0.81\pm0.80$\footnote{This ratio remained constant over the course of our observations.}, which is a typical value for sub-DLAs, but its large error does not allow us to constrain the ionisation parameter $U$ very well. However, because the \ion{H}{i} column density of the GRB\,130606A absorber is only a factor of about two below the DLA definition limit, the ionisation corrections for the elements Fe, Si, and S that we are interested in are moderate regardless of what is assumed for $U$. For example, the sub-DLA in
GRB\,130606A is very similar to the $z=3.142$ sub-DLA in QSO PSS\,J2155+1258 that was presented in Fig.~32 in \citet{2003MNRAS.345..447D}: the neutral hydrogen and
total \ion{Fe}{ii} column densities are practically the same, and the \ion{Si}{ii} column densities are the same within a factor of 2. This analogy suggests that the ionisation correction for the GRB\,130606A sub-DLA for the Fe, S, and Si abundances can
be constrained to $-0.05 \lesssim \epsilon_\mathrm{Fe} < 0$, $-0.2 \lesssim \epsilon_\mathrm{S} < 0$ and $-0.4 \lesssim \epsilon_\mathrm{Si} < 0$, respectively. These limits are independent of the value for $U$ or the strength of the UV radiation field for the calculations considered by \citet{2003MNRAS.345..447D}. Similar constraints on Fe and Si have been inferred for sub-DLAs by \citet{2009MNRAS.397.2037M}; see their Figs. 3 and 4. \citet{2001ApJ...557.1007V} presented a similar study with DLAs, but since our sub-DLA has a column density close to the DLA lower limit, we can extrapolate their model results down to $\log{N_{\ion{H}{i}}}\sim20,$ yielding $\epsilon_\mathrm{Si}\sim-0.21$, $\epsilon_\mathrm{S}\sim-0.30$, $\epsilon_\mathrm{Fe}\sim-0.04,$ and
$\epsilon_\mathrm{Ni}\sim-0.17$. We note that these corrections are negative ($\mathrm{X}^+/\mathrm{H}^0>[\mathrm{X/H}]_\mathrm{total}$), meaning that parts of singly ionised metals are in regions where hydrogen is ionised instead of neutral because the lower column density shields itself less effectively. \\

The ionisation due to the afterglow of GRB\,130606A was calculated with a method similar to the one used in \citet{2012A&A...545A..64D} and \citet{2013A&A...549A..22V}. From published near-IR photometry of GRB\,130606A \citep{2013GCN..14784...1N,2013GCN..14799...1B,2013GCN..14800...1I,2013GCN..14802...1M,2013GCN..14807...1A} and assuming a constant spectral index $\beta=-1$ (see Sect.~\ref{sec:sed}) and a single power-law light curve, we obtain a temporal index $\alpha=-1$, at least until the time at which the spectrum was taken. The absorbing cloud is at distance $d$ from the GRB and is illuminated on one side by the afterglow between $t_0$ and $t_\mathrm{obs}$, the time of our observations. The onset of the afterglow $t_0$ is unknown and introduces an uncertainty; we assumed different values to see its effect. The free parameters are the pre-burst $N_{\ion{H}{i}}$ and $N_{\ion{Fe}{ii}}$. The distance $d$ was obtained from a separate fit using the observed ratio $N_{\ion{Si}{ii}^*}/N_\ion{Si}{ii}$ in the cloud and was estimated to be $d=2.1\pm0.5$, $1.9\pm0.4,$ and $1.8\pm0.4$\,kpc for $t_0=60,\,180,$ and $300$\,s in the rest frame, respectively. These values agree reasonably well with the lower limit $d>2.2\pm0.2$\,kpc derived from the upper limit of $N_{\ion{Fe}{ii}^*}/N_\ion{Fe}{ii}$. The column densities of the ions (except \ion{H}{i} and \ion{Fe}{ii}) are assumed to be equal to their measured values at the start of the simulation (Table~\ref{table:clumns}). This is not entirely correct, but it shows how much they would change when placed at the distance $d$ and were illuminated by the afterglow. From the differences in column density between $t_0$ and $t_\mathrm{obs}$ , we conclude that the ionisation
effects due to the afterglow are very minor. \ion{H}{i} did not change at all, \ion{Fe}{ii} is lowered by $0.03$ dex, \ion{O}{i} by $0.01$, \ion{Si}{ii} by 0.02, \ion{C}{ii} by 0.01, \ion{Al}{ii} by 0.02, and \ion{Al}{iii} by 0.01\,dex by using $t_0=60$\,s and even lower for the higher values of $t_0$ and the larger distance $d>2.2$\,kpc. 

Because our estimates for the ionisation corrections are not the result of fully consistent modelling, we did not apply them directly to our measured abundances. We discuss the estimated metallicity when the corrections are taken into account in Sect. 4.2.

\begin{figure}
\centering
\includegraphics[width=8.4cm,clip=]{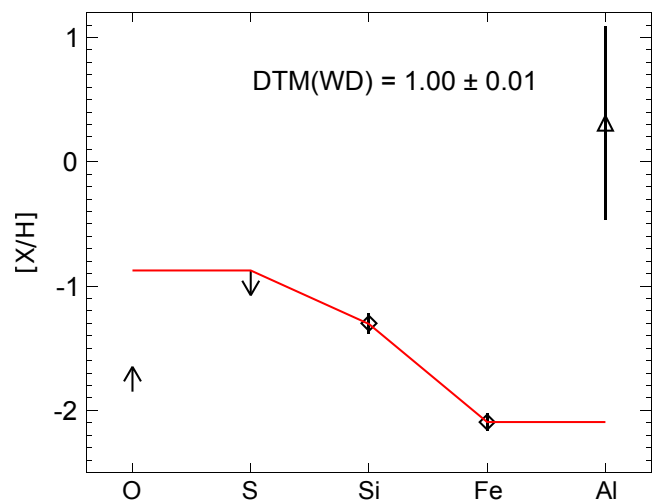}
\caption{Galactic warm disk depletion abundance pattern \citep[solid line,][]{savage1996}, in which Al is assumed to be depleted as strongly as Fe \citep{phillips1982}, and depletion of O is assumed to be negligible \citep{jenkins2009}, fit to the observed pattern (diamonds) following \citet{savaglio2001}. This fit yields $\mathrm{DTM}=1.00\pm0.01$. Aluminium is not included in this fit. The limit on S is taken from \citet{2013arXiv1312.5631C}.}
\label{fig:depletion}
\end{figure}

\subsubsection{Dust depletion}
\label{sec:dustmetal}

Although the optical extinction in the line of sight is low ($A_V<0.2$ mag at the $3\sigma$ level, see Sect.~\ref{sec:sed}), the abundance pattern (Table~\ref{table:clumns}) suggests that there is dust depletion at play. Elements that are locked up into dust by greater amounts (e.g. Si, Fe) have lower column densities in the gas phase than minimally depleted elements (e.g. S) when compared to the relative abundance pattern in the solar photosphere \citep{savage1996,phillips1982,jenkins2009}. 

For the analysis of the dust depletion we assumed that the metallicity is $\mathrm{[S/H]}=-0.88$ (using the limit $\log N_\mathrm{S}<14.17$ from \citealt{2013arXiv1312.5631C}) and included only Si and Fe in the depletion-pattern fit. We followed the method described by \citet{savaglio2001} and refer to depletion levels reported by \citet{savage1996}. We assumed the depletion of oxygen to be negligible \citep{jenkins2009} and the depletion of aluminium to be the same as iron, as suggested in \citet{phillips1982}. The observed abundances are formally best fit by the depletion pattern typical of a warm disk (WD) environment, although the other Galactic depletion patterns \citep{savage1996} cannot be ruled out based on these two measurements. The results are displayed in Fig.~\ref{fig:depletion}. Aluminium is a clear outlier, although the errors on the \ion{Al}{II} column density ($\log N_\ion{Al}{ii}=14.66\pm0.78$) are too large to be constraining. We excluded Al from the fit because it shows an evidently peculiar abundance. Therefore, we cannot compare it to the Galactic depletion patterns of \citet{savage1996}, where solar relative abundances are assumed. We also tested whether an enhancement in the abundance of the $\alpha$-elements could alleviate this discrepancy, but aluminium still remained an outlier. There must be some process that causes an Al enhancement, which we discuss in Sect.~\ref{sec:al}.

The dust-to-metal ratio \citep[DTM, expressed as a fraction of the average Galactic value,][]{Watson11} derived from the depletion-pattern analysis ($\mathrm{DTM} = 1.00 \pm 0.01$) agrees well,  as expected, with the one derived from only the observed [Si/Fe] ($\mathrm{DTM}=1.03 \pm 0.04$), using the method described in \citet{2013A&A...560A..88D}.  A similar DTM was also found when other depletion patterns were used. This means that, assuming the underlying pattern is similar to that of a solar environment (see Sect.~\ref{sec:nucleo}), the dust-to-metal ratio of the host galaxy environment is consistent with the Galactic value, which is remarkable at this redshift. 

\subsection{Intervening absorption systems}
\label{sec:int}
We confirm the detection of several intervening absorbers in
addition to the host galaxy. Based on the preliminary reduction of the same X-shooter spectrum as discussed here, \citet{2013GCN..14816...1X} reported $z_1=2.3103$ (\ion{Mg}{ii}) and $z_3=3.4515$ (\ion{Mg}{ii}, \ion{Fe}{ii}), which we confirm. We additionally detect $z_2=2.5207$ (\ion{Mg}{ii}, \ion{Fe}{ii}), $z_4=4.6448,\,4.6468,\, 4.6495$ (\ion{C}{iv}, \ion{Mg}{ii}, \ion{Al}{ii} \ion{Si}{ii}, \ion{Fe}{ii}) reported earlier by \citet{2013ApJ...774...26C}. We cannot unambiguously confirm the existence of the $z_5=5.806$ system reported by these authors. Lines that are possibly present at this redshift are \ion{Si}{ii}\,$\lambda1260$, \ion{Si}{ii*}\,$\lambda1265$, \ion{O}{i}\,$\lambda1302$, \ion{C}{ii}\,$\lambda1334,$ and \ion{C}{ii*}\,$\lambda1335$, but many strong features are lacking: we do not see any \ion{Fe}{ii} transitions, no \ion{Si}{ii}\,$\lambda1526$, and no \ion{C}{iv}\,$\lambda1548$, although the expected locations of these intrinsically strong lines are in regions with high S/N and only little telluric contamination and should not be blended with the other identified absorbers. 
At the redshifts $z=4.4660,\,4.5309,\,4.5427,$ and $4.6497$ we tentatively detect additional \ion{C}{iv} absorbers, but contamination by other lines and atmospheric features hampers strong conclusions on these systems.

\begin{table}
\caption{Redshift and Doppler parameter of the velocity components fitted to the metal lines of low- and high-ionisation species and for the fit to the Ly$\alpha$ line. $v_{\mathrm{rel}}$ is the velocity relative to that of component 4 (arbitrary).}
\label{table:voigt}
\centering
\begin{tabular}{c c c l l l l l }   
\hline
Component & $z$ & $v_{\mathrm{rel}}$ & $b$ \\
           &                     & (km s$^{-1}$) & (km s$^{-1}$) \\
\hline
Ly$\alpha$ & 5.91248$\pm$0.00028 &$-16$ & 74$\pm$7 \\
\hline
\multicolumn{4}{c}{components of low-ionisation lines}\\
$ 2 $ & $ 5.91182 \pm 0.00002 $ & $-45$ &$ 31 \pm 10 $ \\
$ 4 $ & $ 5.91285 \pm 0.00002 $ & $0$ &$ 14 \pm  2 $\\
$ 5 $ & $ 5.91426 \pm 0.00020 $ & $+61$ &$ 10 \pm  2 $\\
\hline
\multicolumn{4}{c}{components of high-ionisation lines}\\
$ 1 $ & $ 5.91098 \pm 0.00009 $ & $-81$ &$ 37 \pm  9 $ \\
$ 3 $ & $ 5.91265 \pm 0.00003 $ & $-9$ &$ 16 \pm  2 $ \\
$ 6 $ & $ 5.91434 \pm 0.00020 $ & $+64$ &$ 31 \pm 10 $ \\

\hline
\end{tabular}
\end{table}

\begin{table*}
\caption{Derived column densities from the host galaxy absorption lines. The redshift and $b$ parameters of the components for low- and high-ionisation species are listed in Table~\ref{table:voigt}. Column densities are given as $\log(N/\mathrm{cm^{-2}})$.}
\label{table:clumns}
\centering
\begin{tabular}{l c c c l l } 
\hline\hline
\multicolumn{6}{l}{Low-ionisation or neutral species}\\
Ion      & $\log{N_2}$          & $\log{N_4}$           & $\log{N_5}$         & $\log N_\mathrm{tot}$ & [X/H]\\[1px]
\hline
\ion{H}{I} &                               &                            &                       & $19.91\pm0.02$ &   \\
\ion{C}{II} & $ 13.79 \pm 0.14 $ & $ >15.00^a $ & $ 13.83 \pm 0.10 $ & $ >15.04 $ & $[\mathrm{C/H}]>-1.26^{b}$\\ 
\ion{C}{II*} & $ 13.40 \pm 0.15 $ & $ 13.70 \pm 0.09 $ & $ 13.38 \pm 0.11 $ & $ 14.01 \pm 0.06 $ & \\ 
\ion{O}{I} & -- & $ 14.87 \pm 0.12 $ & -- & $ >14.75$ & $[\mathrm{O/H}]>-1.85^{c}$ \\
\ion{Al}{II} & -- & $ 12.72 \pm 0.22 $ & $ 14.63 \pm 0.83 $ & $ 14.66 \pm 0.78 $ & $[\mathrm{Al/H}]=+0.31 \pm 0.78^d$\\ 
\ion{Al}{III} & $ 12.28 \pm 0.54 $ & $ 12.98 \pm 0.14 $ & $ 13.75 \pm 0.23 $ & $ 13.85 \pm 0.19 $ &\\ 
\ion{Si}{II} & $ 12.73 \pm 0.13 $ & $ 13.73 \pm 0.15 $ & $ 13.42 \pm 0.18 $ & $ 13.95 \pm 0.11 $ &  $[\mathrm{Si/H}]=-1.30 \pm 0.08^b$\\ 
\ion{Si}{II*} & $ 12.55 \pm 0.16 $ & $ 13.28 \pm 0.07 $ & $ 13.26 \pm 0.11 $ & $ 13.62 \pm 0.06 $ &\\ 
\ion{S}{II} & $ 13.55 \pm 0.31 $ & $ 12.95 \pm 1.04 $ & $ 13.02 \pm 0.74 $ & $<14.44^e$ &$ [\mathrm{S/H}]<-0.60$\\ 
\ion{Fe}{II} & -- & $ 13.09 \pm 0.05 $ & $ 12.82 \pm 0.17 $ & $ 13.29 \pm 0.07 $ & $[\mathrm{Fe/H}]=-2.09 \pm 0.08$\\
\ion{Fe}{II*} & $<11.80$ & $<11.80$ & $<11.80$ & $<12.10$&\\ 
\ion{Ni}{II} & -- & $ <13.83$ & $ <13.91$ & $<14.22$ & $[\mathrm{Ni/H}]<+0.16^b$\\ 
\ion{Ni}{II*} & $<12.92$ & $<12.90$ & $<13.14$ & $<13.45 $ &\\ 
\hline
\multicolumn{6}{l}{High-ionisation species}\\
Ion     & $\log{N_1}$           & $\log{N_3}$           & $\log{N_6}$         & $\log N_\mathrm{tot}$ & \\
\hline
\ion{N}{v} & $ 13.79 \pm  0.06$ & $ 14.04\pm  0.09 $ & $ 14.33  \pm   0.04$ & $14.59 \pm 0.03 $ & \\ 
\hline
\end{tabular}
\tablefoot{$^a$This component is most likely saturated; we obtain a minimum $\log N$ from the equivalent width of the component.\\
$^b$Based on the sum of the ground state and excited state.\\
$^c$Lower limit because both component 5 and the excited state column density cannot be measured due to telluric contamination.\\
$^d$Based only on \ion{Al}{ii}. The sum of \ion{Al}{ii} and \ion{Al}{iii} results in $[\mathrm{Al/H}]=+0.46 \pm 0.62$.\\
$^e$$1\sigma$ upper limit because the relative strength of the components does not resemble the clearly detected lines, which
means that we are measuring the noise.}
\end{table*}

\subsection{Ly$\alpha$ forest constraints on the IGM}
\label{sec:IGM}
In this section we analyse the ionisation state of the IGM at $z\sim5-6$ via the Gunn-Peterson optical depth \citep{1965ApJ...142.1633G}.
\subsubsection{Ly$\alpha$ absorption}

We measured the Ly$\alpha$ absorption by the IGM from $1200\,\AA$ in the rest frame, which is the limit not affected by the blue wing of Ly$\alpha$, to $1040\,\AA$, the shortest wavelength that is not affected by possible Ly$\beta$ emission from the GRB host. Within this range, we used redshift intervals of $\Delta z$ of 0.15, which corresponds to $\sim$60\,Mpc in comoving distance, for comparison purposes with previous work. Using this redshift range and wavelength interval, we measured the transmission as 
\begin{equation}
\mathcal{T}(z_\mathrm{abs})\equiv \langle f_{\nu}^\mathrm{obs}/f_{\nu}^\mathrm{int}\rangle.
\label{trans}
\end{equation}
The measured transmitted flux ($\mathcal{T}$) is shown in Table~\ref{tab:transtau} and plotted in Fig.~\ref{fig:trans}. The continuum level is determined first by fitting a power law to the near-infrared part of the spectrum, taking advantage of the broad wavelength coverage of X-shooter. Then, we fixed the slope and fitted the normalisation to the optical part of the spectrum ($8530-8950\,\AA$). We masked strong identified absorption features (both telluric and extragalactic) in this process. When using QSOs for this type of analysis, continuum determination is often the largest source of uncertainty because
of the complicated shape of their spectrum. 

A GRB continuum can be more accurately predicted as a result
of its relatively flat shape that shows no broad features. 
We estimated errors on the transmission by changing the continuum slope within 1$\sigma$ (see Sect.~\ref{sec:sed}), added in quadrature to the noise in the spectra and the errors in determining the unabsorbed continuum level. The latter dominates the errors on $\mathcal{T}$.

It is conventional to express the optical depth $\tau$ in terms of $\tau=-\ln(\mathcal{T})$. We present $\tau$ as a function of redshift in Fig.~\ref{fig:tau}. 

\subsubsection{Ly$\beta$ absorption}

At the same neutral hydrogen density, the optical depth $\tau$ is proportional to $f\lambda_0$, where $f$ and $\lambda_0$ are the oscillator strength and rest-frame wavelength of the transition, respectively. Therefore, the optical depth of Ly$\beta$ is a factor of 6.2 smaller than that of Ly$\alpha$ in a homogeneous medium illuminated by a uniform radiation field. This means that Ly$\beta$ probes into a larger amount of neutral hydrogen than Ly$\alpha$. In this section we measure the optical depth using Ly$\beta$ absorption from the spectra. 

We assumed the same continuum as we used for Ly$\alpha$. We chose the minimum wavelength to be $970\,\AA$, above which it is not affected by Ly$\gamma$ absorption. This results in a redshift range of $5.59<z_\mathrm{abs}<5.74$ for Ly$\beta$. The Ly$\beta$ absorptions overlap with Ly$\alpha$ absorption at lower redshift. Therefore, we corrected the Ly$\alpha$ absorption to measure $\tau_{\beta}$. We used Eq. 5 of \citet[][  ]{2006AJ....132..117F} to estimate Ly$\alpha$ absorption from lower redshift and corrected the Ly$\beta$ transmission measurements. Table~\ref{tab:transtau} shows the results, which are graphically represented in Fig.~\ref{fig:trans} as blue diamonds.

To convert Ly$\beta$ transmission to $\tau_{\alpha}$, different optical depths between Ly$\alpha$ and Ly$\beta$
have to be considered. The difference depends on the UV background and its uniformity, the clumpiness of the IGM, and its equation of state. According to simulations \citep[e.g.,][]{2005ApJ...620L...9O} and empirical measurements \citep{2006AJ....132..117F}, the $\tau_{\alpha}/\tau_{\beta}$ conversion is in the range of $2.2-2.9$.  The $\tau_{\alpha}/\tau_{\gamma}$ conversion lies in the range of $4.4-5.7$.
Following the discussion in \citet{2006AJ....132..117F}, we used  $\tau_{\alpha}/\tau_{\beta}=2.25$ and
$\tau_{\alpha}/\tau_{\gamma}=4.4$. These values are based on measurements of transmitted flux at $5.4<z<5.8$ in the 
QSO sample of \citet{2006AJ....132..117F}.
 
Figure~\ref{fig:tau} shows constraints on the $\tau_{\alpha}$ from Ly$\beta$ absorption measurements. We note that $\tau$ is converted to the Ly$\alpha$ optical depth in Fig.~\ref{fig:tau}. 

\subsubsection{Ly$\gamma$ absorption}
Similar to Ly$\beta$, the Ly$\gamma$ optical depth ($f\lambda_0$) is a factor of 17.9 lower than that of Ly$\alpha$, providing us with a chance to probe into even more neutral hydrogen. The more neutral regions are denser in general. Therefore, Ly$\gamma$ transmission also provides the opportunity to probe regions with different density, potentially constraining the density-temperature relation.

The Ly$\gamma$ absorption measurement is restricted by Ly$\beta$ absorption at lower redshifts and the Ly$\delta$ absorption at the higher redshifts (closer to the GRB). Therefore, we used a smaller bin size of $\Delta z= 0.06$ than for the Ly$\alpha$ and Ly$\beta$ transmission. The Ly$\gamma$ transmission measurements were corrected for overlapping lower redshift Ly$\alpha$ and Ly$\beta$ absorptions using the best-fit power laws to lower redshift data \citep[][their Eqs. 5 and 6]{2006AJ....132..117F}.
The optical depth was converted to $\tau_{\alpha}$, using $\tau_{\alpha}/\tau_{\gamma}=4.4$.  The results are listed in Table~\ref{tab:transtau} and shown in Fig.~\ref{fig:tau}. 

The Ly$\gamma$ absorption measurements are much more challenging than those of Ly$\alpha$ and Ly$\beta$. Since the overlapping foreground Ly$\alpha$ and Ly$\beta$ lines absorb $\sim$98\% of the continuum flux in the wavelength ranges of Ly$\gamma$, we need to measure absorption in the remaining $\sim$2\%. Errors in the continuum determination are also larger in this region since we use the bluer part of the spectrum.\\

\begin{figure}
\centering
\includegraphics[width=8.2cm]{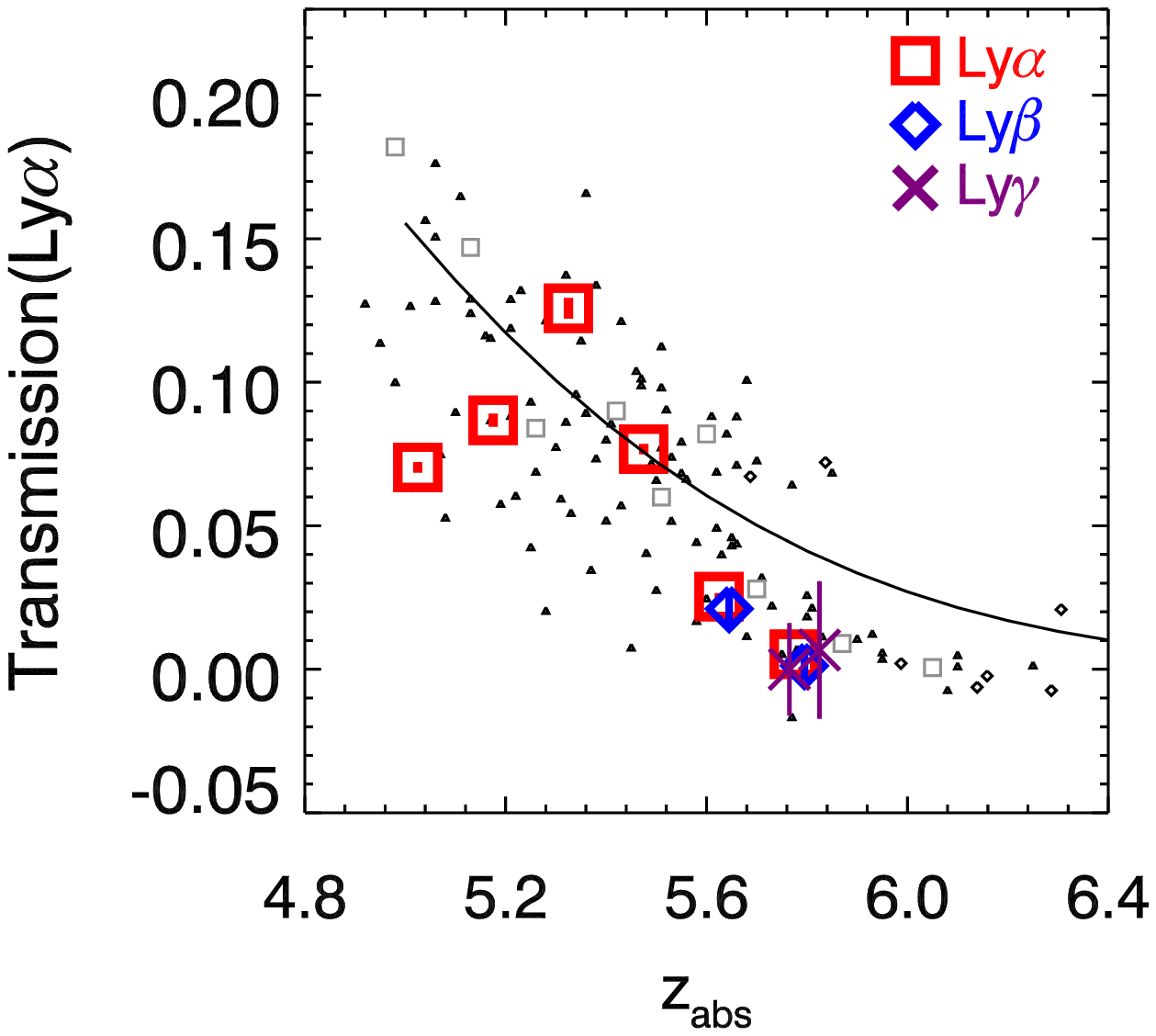}
\caption{
Ly$\alpha$ transmission in the spectrum of GRB\,130606A (red squares). Ly$\beta$ (blue diamonds) and Ly$\gamma$ (purple crosses) measurements are converted to equivalent values at Ly$\alpha$. The black triangles, diamonds, and the grey squares are measurements based on quasars from \citet{2006AJ....132..117F}, \citet{2011MNRAS.415L...1G}, and \citet{2004AJ....127.2598S}, respectively. The solid line is the best power-law fit to the data at $z<5.5$ by \citet[][their Eq. 5]{2006AJ....132..117F}. The uncertainties of the QSO points are typically a factor of two. 
}\label{fig:trans}
\end{figure}

\begin{figure}
\centering
\includegraphics[width=8.2cm]{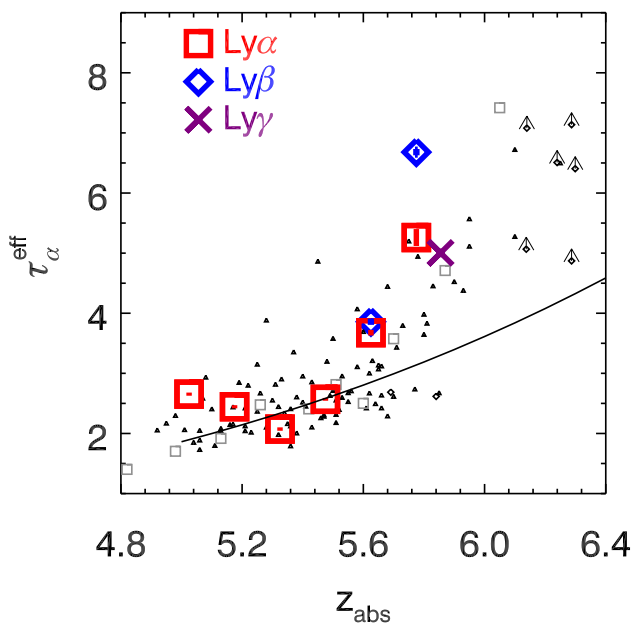}
\caption{Effective Gunn-Peterson Ly$\alpha$ optical depth of the IGM from the spectrum of
\thisgrb. The red squares, blue diamonds, and purple crosses are measurements with error estimates
from Ly$\alpha$, Ly$\beta,$ and Ly$\gamma$, respectively, the latter two converted to the Ly$\alpha$ optical depth. The black triangles, diamonds, and the grey squares are measurements based on QSO spectra from \citet{2006AJ....132..117F}, \citet{2011MNRAS.415L...1G}, and \citet{2004AJ....127.2598S}, respectively. The uncertainties of the QSO points are typically a factor of two. The solid line is the best power-law fit to the data at $z<5.5$ by \citet[][their Eq. 5]{2006AJ....132..117F}.
}\label{fig:tau}
\end{figure}

In Figs.~\ref{fig:trans} and \ref{fig:tau}, small triangles and grey squares represent previous data based on quasars from \citet{2006AJ....132..117F} and \citet{2004AJ....127.2598S}.  The solid line shows the best power-law fit to the data at $z<5.5$ based on \citet[][their Eq. 5]{2006AJ....132..117F}. Within the scatter, our measurements agree with earlier findings. At $z>5.7$, our data points deviate from the fit to $z<5.5$ data, suggesting that the Universe was not yet completely ionised at $z>5.7$.

\begin{table}
\begin{center}
\caption{IGM absorption  towards the GRB\,130606A at $z=5.913$.  The bin widths are  $\Delta z=0.15,0.15$, and 0.06 for Ly$\alpha$, Ly$\beta$, and Ly$\gamma$~  absorptions, respectively.}
\label{tab:transtau}
\begin{tabular}{ccrl}
  \hline
   Redshift & Line & Transmission & $\tau_{\alpha}$  \\ 
 \hline
5.77  & Ly$\alpha$ & 0.0052$\pm$0.001 &   5.26$\pm$   0.14\\
5.62  & Ly$\alpha$ & 0.0253$\pm$0.001 &   3.68$\pm$   0.05\\
5.47  & Ly$\alpha$ & 0.0767$\pm$0.002 &   2.57$\pm$   0.02\\
5.33  & Ly$\alpha$ & 0.1258$\pm$0.004 &   2.07$\pm$   0.03\\
5.18  & Ly$\alpha$ & 0.0868$\pm$0.002 &   2.44$\pm$   0.03\\
5.02  & Ly$\alpha$ & 0.0704$\pm$0.002 &   2.65$\pm$   0.03\\
5.78  & Ly$\beta$ &  0.051$\pm$0.005 &   6.68$\pm$   0.09\\
5.62  & Ly$\beta$ &  0.180$\pm$0.006 &   3.86$\pm$   0.03\\
5.84  & Ly$\gamma$ &   0.32$\pm$0.024 &    5.0$\pm$    0.1\\
5.79  & Ly$\gamma$ &   0.08$\pm$0.016 &   18.2$\pm$    0.2\\
\hline
\end{tabular}
\end{center}
\end{table}

\section{Discussion}

\subsection{Analysis of the red damping wing}
\label{sec:disc:redwing}
\begin{figure}
\centering
\includegraphics[width=8.6cm]{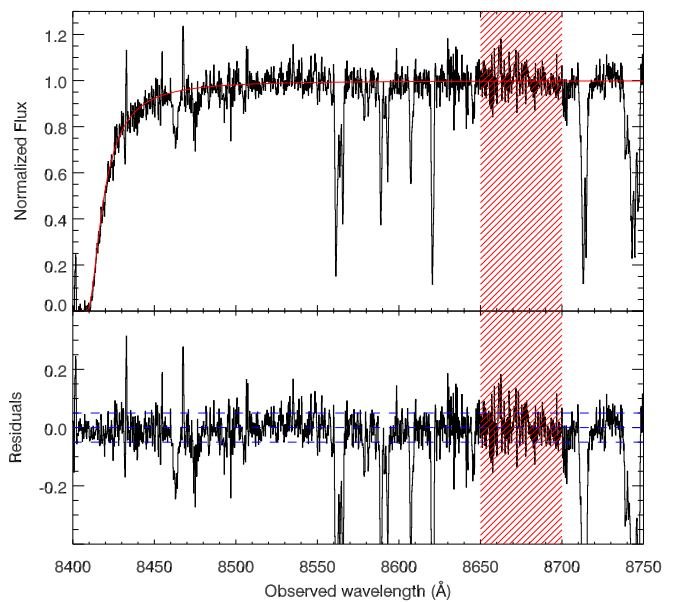}
\caption{Excerpt of the X-shooter spectrum with the same model fit as shown in Fig.~\ref{lyafig}. The lower panel shows the residuals and deviations from the fit. The red hatched region is range III in \citet{2014PASJ...66...63T}, where their spectrum is inconsistent with a single DLA model. With our data, we can obtain a good fit with this model. The difference is most likely due to a different spectral index $\beta$, see Sect.~\ref{sec:disc:redwing}.
}\label{fig:totani}
\end{figure}

\citet{2014PASJ...66...63T} reported evidence for a high neutral fraction based on their spectrum of the afterglow. In particular, they found that a Voigt-profile fitted to the red wing of the absorption trough is inconsistent with the data in a region referred to as wavelength range III by these authors, from about $8650$ to $8700\,{\AA}$. In Fig.~\ref{fig:totani} we show this wavelength range III in our spectrum including our fit and the deviations from the fit. Our spectrum is fully consistent with the Voigt-profile fit without a neutral hydrogen IGM component (see also Sect.~\ref{sec:reddamping}). One important difference between our analysis and that of \citet{2014PASJ...66...63T} is that
the value of the spectral slope adopted by these authors ($\beta = -0.74$) is inconsistent with the one we find. 

We subsequently discussed this with Dr Totani and collaborators and made our spectrum available to them. The difference between the obtained neutral fractions seems to some extent to be dependent on the fitting method and on which sections of the spectrum are fitted. Dr Totani and collaborators will carry out an independent analysis of our spectrum (Totani et al., in preparation).

\subsection{Metallicity of the host of \thisgrb}
\label{sec:disc:metabun}
The `raw' measured column density values and limits we derived for the host-galaxy ISM (Table~\ref{table:clumns}) agree well with the results from \citet{2013ApJ...774...26C} and \citet{2013arXiv1312.5631C}. Because our data have a higher spectral resolution and a larger wavelength coverage and because all results of other observations have been published, we try to explain the effects of ionisation and dust depletion to better constrain metallicity and true abundance pattern in the host galaxy of GRB\,130606A. Estimates on the effects of ionisation are provided in Sect.~\ref{sec:ioncor}, where we conservatively assumed the following contraints on the ionisation corrections, including both pre-burst and post-burst effects: $\epsilon_\mathrm{Fe} \sim 0$, $-0.4 \lesssim \epsilon_\mathrm{S} < 0$ and $-0.4 \lesssim \epsilon_\mathrm{Si}<0$. Here, we discuss their possible implications. We measured $[\mathrm{S/H}]<-0.60$, while the spectrum by \citet{2013arXiv1312.5631C}, taken earlier in time, provides are more stringent upper limit of $\log N_\mathrm{S}<14.17$ and $[\mathrm{S/H}]<-0.88$. Sulphur is only minimally depleted onto dust, and considering that the sulphur ionisation correction would only make this abundance smaller, this value should reflect a real upper limit on the metallicity. For silicon, we measured $[\mathrm{Si/H}]=-1.30\pm0.08$; adding the strongest ionisation correction, this values becomes $[\mathrm{Si/H}]_\mathrm{ic}=-1.7$, which is a stringent lower limit on the metallicity. From the dust depletion analysis we know that a fraction of the Si is in dust, but the analysis is not robust enough to estimate an
exact value. The effect is included, while Si provides the lower limit. The iron abundance $[\mathrm{Fe/H}]=-2.09\pm0.08$ is negligibly affected by ionisation, but severely so by dust depletion and does not provide additional information on the metallicity. The metallicity is constrained to be $-1.7<[\mathrm{M/H}]<-0.9$, since Si provides a lower limit and S the upper limit. This is relatively high when compared to QSO-DLAs at $z>5$ \citep{2012ApJ...755...89R,2014ApJ...782L..29R}, but not unexpected given predictions in recent models \citep{2013MNRAS.429.2718S}. The difference in metallicity evolution in GRB-DLAs and QSO-DLAs has also recently been discussed in \citet{2013MNRAS.428.3590T,2014ApJ...785..150S}, and
\citet{2014arXiv1408.3578C}, for instance.

\subsection{Nucleosynthetic history}
\label{sec:nucleo}
The high value of [Si/Fe] can be fully attributed to dust depletion, but the effects of this, and those of an underlying pattern that is different from that in the solar photosphere (e.g., $\alpha$-element enhancement), are degenerate. At this redshift, one would expect to see more elements in the ISM created by shorter-lived massive stars, $\alpha$-elements (C, N, O, Ne, Mg, S, Si, Ar, Ca, Ti) from core-collapse supernovae (SNe) instead of products of SNe originating in longer-living stars: the iron peak elements created by type Ia SNe (V, Cr, Mn, Fe, Co, Ni). This has not unambiguously been detected in high-redshift GRB host galaxies to date because there usually is the combined effect of the dust depletion \citep[see, e.g.,][]{2013MNRAS.428.3590T,2014ApJ...785..150S}. With the limited information we have for GRB\,130606A, we cannot draw strong conclusions on either the dust depletion or $\alpha$-element enhancement. With a pure dust-depletion explanation, the dust-to-metal ratio is about the same as the Galactic value. The dust extinction ($A_V < 0.2\,\rm{mag}$ at $3\sigma$) can still be low due to the low total metal column density \citep[see also, e.g.,][]{2011A&A...532A.143Z}. A peculiarity that we cannot explain with dust depletion is the very high abundance of aluminium, which we discuss in Sect.~\ref{sec:al}. 

\subsubsection{High aluminium abundance at $z\sim6$:  a relic of Population~III stars?}
\label{sec:al}
The analysis of the X-shooter spectrum of the host galaxy of \thisgrb~yields a remarkably high aluminium abundance: $[\ion{Al}{ii}/\ion{H}{i}] = +0.31 \pm 0.78$, 
given the low iron abundance of $[\mathrm{Fe}/\mathrm{H}] = -2.09 \pm 0.08$. While the relatively large errors on the aluminium abundance cause this observation to be only marginally significant ($\approx 3\sigma$), it is not the only unusual abundance ratio in this system.
The oxygen abundance is very low: \citet{2013ApJ...774...26C} found $[\mathrm{O}/\mathrm{H}]  \sim-2.0$, and we derive a lower limit of $[\mathrm{O}/\mathrm{H}]  > -1.85$,
while the silicon abundance, $[\mathrm{Si}/\mathrm{H}]  = -1.30\pm0.08$, is much higher than the iron abundance.

This unusual abundance pattern is not observed for the first time in a GRB host. In the host galaxy of GRB\,120327A \citep[$z = 2.8145$,][]{delia2014} the aluminium abundance was also high $[\mathrm{Al}/\mathrm{H}]  = 0.00\pm0.11$; $\mathrm{[Al/Fe]}=1.73\pm0.07$), although saturation of the \ion{Al}{ii} and \ion{Al}{iii} lines prevents an accurate abundance analysis of the three individual absorption line components. The oxygen abundance ($[\mathrm{O}/\mathrm{H}] = -1.98\pm0.13$) is also lower than that of iron ($[\mathrm{Fe}/\mathrm{H}] = -1.73\pm0.10$), and the silicon abundance is again high ($[\mathrm{Si}/\mathrm{H}]  = -1.16\pm0.09$).
The similarity between these two unusual patterns deserves some discussion, and we here consider possible causes.

One explanation might be ionisation corrections due to the fairly low column density of \thisgrb, but as shown in Sect.~\ref{sec:ioncor}, this correction is less than 0.4 dex for \thisgrb\footnote{Larger ionization corrections for [Al/Fe] are supported by the calculations
of \citet{2001ApJ...557.1007V}, with corrections potentially as large as $\sim1.4$ dex. If we were to consider this (much larger) ionisation correction, we would still conclude that \thisgrb\,exhibits an unusually large enhancement in [Al/Fe].}. Furthermore, the host of GRB\,120327A has an $\ion{H}{i}$ column density of $\log(N_{\ion{H}{i}}/\mathrm{cm}^{-2})=22.01$ and therefore will need no ionisation corrections. Taken together, it is very unlikely that this is the cause.

Another possibility is depletion onto dust grains. Our understanding of dust depletion is still very far from complete, and much is yet to be learned in this field. However, the depletion of Al and Fe is is expected to be very similar \citep{phillips1982}, so again while hypothetically possible, we consider this an unlikely scenario.

An interesting, and possibly related, observation is the relatively high aluminium abundance that has been measured in the atmospheres of red giants in metal-poor
($[\mathrm{Fe}/\mathrm{H}]  < -1$) globular clusters \citep[e.g.,][]{1997AJ....113..279K}. In these atmospheres, [Al/Fe] seems to be always $\lesssim1.5$ \citep{2014ApJ...780...94C}, consistent with our value of $+2.40\pm0.78$, as well as that of GRB\,120327A ($+1.73\pm0.07$), although both of the GRB host measurements are in
the upper range. In these globular clusters a strong anti-correlation has been found between the oxygen and sodium abundance. This anti-correlation has been interpreted as being due to proton-capture nucleosynthesis, where Na and Al are produced at the expense of Ne and Mg in regions where C and O are converted into N, that
is, where the CNO
cycle is active \citep{1998ApJ...492..575C}. Likewise, Si would be produced by proton capture of Al. Regarding these metal-poor globular clusters, the question is whether the proton capture occurred within the observed red giants, that is, during shell burning when the temperature is high enough for the CNO cycle, and convection dredged up the metals produced in these deeper layers within the stars (evolutionary scenario). Alternatively, these abundance ``anomalies" might have been produced by proton capture in a previous generation of massive (perhaps even the first) stars (primordial scenario). The detection of similar abundance anomalies in main-sequence, turn-off and early sub-giant stars \citep{2001A&A...369...87G} suggests the latter scenario. \citet{2014MNRAS.437L..21D} proposed that the abundance anomalies of proton-capture elements in globular
clusters were produced by supermassive stars with $M\sim10^4\,\mathrm{M}_\odot$, products of the runaway collisions of massive stars in dense clusters \citep{2004Natur.428..724P}.

It appears therefore that the measured abundances in the two GRB host galaxies discussed here find their simplest and most logical explanation as the result of
proton-capture at work. Given the high redshift ($z=5.913$) of \thisgrb and the corresponding look-back time ($\sim12.6\,\mathrm{Gyr}$), these observations are consistent with the scenario that a previous generation of massive stars must have produced the chemical enrichment via proton capture.

\subsection{Reionisation of the IGM}
Prompt follow-up spectroscopy of \thisgrb~has enabled to reliably measure Gunn-Peterson optical depths using a GRB instead of a QSO as a background source \citep[see also][]{2013ApJ...774...26C,2014PASJ...66...63T}. The only other GRB for which this has been attempted is GRB\,050904 \citep{2006PASJ...58..485T}. Given a high S/N afterglow spectrum, GRBs have several advantages over QSOs; the simple power-law continuum of a GRB spectrum is much easier to estimate than the complicated shape of a QSO spectrum, including a possibly broken power-law UV continuum, whose break is in the heavily absorbed UV range at high-redshift and broad emission lines.
As a consequence of this, the uncertainties of the measured optical depths based on our GRB spectrum are much smaller than typical uncertainties on values in previous work based on QSOs, which are about a factor of 2. 
In addition, GRBs can be much brighter than QSOs if observed early enough. At the time of our reported spectroscopy, \thisgrb~had a near-IR brightness ($Y,\,J,\,H$) of around the $18^\mathrm{th}$ magnitude \citep{2013GCN..14799...1B}. This is much brighter than $z\sim6$ QSOs used in previous work ($m_z\sim20$\,mag). While obtaining high S/N spectra of fainter QSOs with current 8\,m class telescopes becomes progressively more difficult at higher redshift with fainter QSOs, we can in principle extend the Gunn-Peterson test to much higher redshifts by using
GRBs, without having to rely on future larger optical telescopes. 
 
Another advantage of GRBs over QSOs is that the sightlines towards them suffer from different biases. GRB host galaxies differ from typical field galaxies \citep{2006Natur.441..463F,2009ApJ...691..182S,2014ApJS..213...15W}. At low redshift they are found to be low-mass, metal-poor galaxies, which reside in representative regions of the Universe. On the other hand, QSOs tend to be contained in massive haloes. The surroundings of QSOs could be overdense on scales of several tens of Mpc \citep[for example see][]{2009MNRAS.394..577O,2010ApJ...721.1680U}. In this case, the IGM in front of QSOs could be more ionised than a typical region of the Universe. Furthermore, the QSO background light has been present for a long time, which may have affected the degree of ionisation of the material in the sight line, in particular close to the source: the proximity effect. This is not the case for GRB afterglows.

However, despite our accurate measurements, the optical depths show significant variation across the redshift window we used ($5.0<z<5.8$). The variation is much larger than our measurement errors, and therefore, intrinsic. This suggests a strong spatial variation in optical depths at this epoch of the Universe. To fully understand the reionisation history of the Universe, it is therefore important to expand the statistical sample. GRBs will probably play a prominent role in this.

\section{Conclusions}
We reported the analysis of the $z=5.913$ GRB\,130606A afterglow spectrum obtained with VLT/X-shooter. The analysis can be divided into two main parts: the study of the abundance pattern in the host galaxy measured by the afterglow absorption lines, and the study of the ionised fraction of the IGM through the shape of the red wing of Ly$\alpha$ at $z_\mathrm{GRB}$ and through the Gunn-Peterson optical depth in front of the GRB ($5.02<z<5.84$).

Although many host absorption lines were detected, the abundances
can only be well constrained for H, Si, Fe, and Al; for C, O, S, and Ni we obtained limits. The high $\mathrm{[Si/Fe]}=+0.79\pm0.13$ can be explained with dust depletion with a dust-to-metal ratio similar to the Galactic value or $\alpha$-element enhancement. The abundance of aluminium is very high: $\mathrm{[Al/H]}=+0.31\pm0.78$ and $\mathrm{[Al/Fe]}=+2.40\pm0.78$ based on \ion{Al}{ii} alone, and even higher when taking into account \ion{Al}{iii}, which shows a similar line structure, suggesting that the region where this resides is associated with the region in which \ion{Al}{ii} is located. With $\log N_\ion{H}{i} =19.91\pm0.02$ the absorber is a sub-DLA and might not be as efficiently shielding the ions from ionisation as DLAs. We estimated the ionisation corrections both in the pre-burst sub-DLA and due to the UV radiation GRB afterglow. While the former results in corrections of up to 0.4\,dex, the latter is in this case negligible with corrections below $0.03$\,dex. When these corrections and the more stringent upper limit on S by \citet{2013arXiv1312.5631C}
are taken into account, the metallicity is estimated to be $-1.7<[\mathrm{M/H}]<-0.9$ ($2\%-13\%$ of solar). The metallicity and low-ionisation line width measured at this high redshift are consistent with the break in the evolution of the mass-metallicity relation for DLAs proposed in \citet{2013MNRAS.430.2680M}, but also marginally with a model that assumes a shallower slope and no break \citep{2013ApJ...769...54N}.

From fitting the red wing of the Ly$\alpha$ absorption line of the GRB host galaxy, we concluded that the IGM is predominantly neutral: the ionised fraction $x_{\ion{H}{i}} = 0$ ($x_{\ion{H}{i}} < 0.03$ at 3$\sigma$ significance). We measured the effective Gunn-Peterson Ly$\alpha$ optical depth of the IGM at $5.02<z<5.84$. Our well-constrained data points agree with earlier test with background QSOs, which showed that the IGM was increasingly neutral at $z>5.6$ but still overwhelmingly ionised. However, the intrinsic scatter within the measurement is much larger than the errors on the individual data points, which most likely reflects that the reionisation is a complicated process with a strong spatial variation. A larger statistical sample is required to understand the IGM state at the end of reionisation.

\begin{acknowledgements}
We thank the anonymous referee for careful inspection of the manuscript and constructive comments. We thank Max Pettini, Darach Watson, Georges Meynet, and Tomonori Totani for helpful discussions. We thank Andrea Rossi for careful reading of the manuscript. We thank Maryam Arabsalmani, who re-ran her code for us including our data point in Sect.~\ref{sec:relstrengths}.
O.\,E.\,H acknowledges the Dutch Research School for Astronomy (NOVA) for a PhD grant.
D.\,M acknowledges the Instrument center for Danish Astrophysics (IDA) for funding. The research leading to these results has received funding from the European Research Council under the European Union's Seventh Framework Program (FP7/2007-2013)/ERC Grant agreement no. EGGS-278202 (J.\,P.\,U.\,F).  T.\,K
acknowledges support by the European Commission under the Marie Curie Intra-European Fellowship Programme in FP7. The research of A.d.U.P. is supported by the Spanish project AYA2012-39362-C02-02 and by the European Commission under the Marie Curie Career Integration Grant programme (FP7-PEOPLE-2012-CIG 322307). The Dark Cosmology Centre is funded by the DNRF.

\end{acknowledgements}

\bibliographystyle{aa}
\bibliography{ref}

\begin{thebibliography}{109}
\expandafter\ifx\csname natexlab\endcsname\relax\def\natexlab#1{#1}\fi

\bibitem[{{Afonso} {et~al.}(2013){Afonso}, {Kann}, {Nicuesa Guelbenzu},
  {Kruehler}, {Elliott}, \& {Greiner}}]{2013GCN..14807...1A}
{Afonso}, P., {Kann}, D.~A., {Nicuesa Guelbenzu}, A., {et~al.} 2013, GRB
  Coordinates Network, 14807

\bibitem[{{Andersen} {et~al.}(2000){Andersen}, {Hjorth}, {Pedersen}, {Jensen},
  {Hunt}, {Gorosabel}, {M{\o}ller}, {Fynbo}, {Kippen}, {Thomsen}, {Olsen},
  {Christensen}, {Vestergaard}, {Masetti}, {Palazzi}, {Hurley}, {Cline},
  {Kaper}, \& {Jaunsen}}]{2000A&A...364L..54A}
{Andersen}, M.~I., {Hjorth}, J., {Pedersen}, H., {et~al.} 2000, \aap, 364, L54

\bibitem[{{Arabsalmani} {et~al.}(2015){Arabsalmani}, {M{\o}ller}, {Fynbo},
  {Christensen}, {Freudling}, {Savaglio}, \& {Zafar}}]{2015MNRAS.446..990A}
{Arabsalmani}, M., {M{\o}ller}, P., {Fynbo}, J.~P.~U., {et~al.} 2015, \mnras,
  446, 990

\bibitem[{{Asplund} {et~al.}(2009){Asplund}, {Grevesse}, {Sauval}, \&
  {Scott}}]{2009ARA&A..47..481A}
{Asplund}, M., {Grevesse}, N., {Sauval}, A.~J., \& {Scott}, P. 2009, \araa, 47,
  481

\bibitem[{{Barthelmy} {et~al.}(2013){Barthelmy}, {Baumgartner}, {Cummings},
  {Fenimore}, {Gehrels}, {Krimm}, {Lien}, {Markwardt}, {Palmer}, {Sakamoto},
  {Sato}, {Stamatikos}, {Tueller}, \& {Ukwatta}}]{2013GCN..14819...1B}
{Barthelmy}, S.~D., {Baumgartner}, W.~H., {Cummings}, J.~R., {et~al.} 2013, GRB
  Coordinates Network, 14819

\bibitem[{{Berger} {et~al.}(2007){Berger}, {Chary}, {Cowie}, {Price},
  {Schmidt}, {Fox}, {Cenko}, {Djorgovski}, {Soderberg}, {Kulkarni}, {McCarthy},
  {Gladders}, {Peterson}, \& {Barger}}]{2007ApJ...665..102B}
{Berger}, E., {Chary}, R., {Cowie}, L.~L., {et~al.} 2007, \apj, 665, 102

\bibitem[{{Butler} {et~al.}(2013){Butler}, {Watson}, {Kutyrev}, {Lee},
  {Richer}, {Klein}, {Fox}, {Prochaska}, {Bloom}, {Cucchiara}, {Troja},
  {Littlejohns}, {Ramirez-Ruiz}, {de Diego}, {Georgiev}, {Gonzalez},
  {Roman-Zuniga}, {Gehrels}, \& {Moseley}}]{2013GCN..14799...1B}
{Butler}, N., {Watson}, A.~M., {Kutyrev}, A., {et~al.} 2013, GRB Coordinates
  Network, 14799

\bibitem[{{Castro-Tirado} {et~al.}(2013{\natexlab{a}}){Castro-Tirado},
  {S{\'a}nchez-Ram{\'{\i}}rez}, {Ellison}, {Jel{\'{\i}}nek},
  {Mart{\'{\i}}n-Carrillo}, {Bromm}, {Gorosabel}, {Bremer}, {Winters},
  {Hanlon}, {Meegan}, {Topinka}, {Pandey}, {Guziy}, {Jeong}, {Sonbas},
  {Pozanenko}, {Cunniffe}, {Fern{\'a}ndez-Mu{\~n}oz}, {Ferrero}, {Gehrels},
  {Hudec}, {Kub{\'a}nek}, {Lara-Gil}, {Mu{\~n}oz-Mart{\'{\i}}nez},
  {P{\'e}rez-Ram{\'{\i}}rez}, {{\v S}trobl}, {{\'A}lvarez-Iglesias},
  {Inasaridze}, {Rumyantsev}, {Volnova}, {Hellmich}, {Mottola}, {Castro
  Cer{\'o}n}, {Cepa}, {G{\"o}{\u g}{\"u}{\c s}}, {G{\"u}ver}, {{\"O}nal Ta{\c
  s}}, {Park}, {Sabau-Graziati}, \& {Tejero}}]{2013arXiv1312.5631C}
{Castro-Tirado}, A.~J., {S{\'a}nchez-Ram{\'{\i}}rez}, R., {Ellison}, S.~L.,
  {et~al.} 2013{\natexlab{a}}, ArXiv e-prints, 1312.5631

\bibitem[{{Castro-Tirado} {et~al.}(2013{\natexlab{b}}){Castro-Tirado},
  {Sanchez-Ramirez}, {Gorosabel}, {Jelinek}, {Tello}, {Ferrero}, {Lara-Gil},
  {Cunniffe}, {Perez-Ramirez}, {Kubanek}, {Castro Ceron}, {Mottola},
  {Hellmich}, {Fernandez-Munoz}, {Munoz-Martinez}, {Sabau-Graziati},
  {Martin-Carrillo}, {Cepa}, {Tejero}, \&
  {Alvarez-Iglesias}}]{2013GCN..14796...1C}
{Castro-Tirado}, A.~J., {Sanchez-Ramirez}, R., {Gorosabel}, J., {et~al.}
  2013{\natexlab{b}}, GRB Coordinates Network, 14796

\bibitem[{{Cavallo} {et~al.}(1998){Cavallo}, {Sweigart}, \&
  {Bell}}]{1998ApJ...492..575C}
{Cavallo}, R.~M., {Sweigart}, A.~V., \& {Bell}, R.~A. 1998, \apj, 492, 575

\bibitem[{{Chen} {et~al.}(2009){Chen}, {Perley}, {Pollack}, {Prochaska},
  {Bloom}, {Dessauges-Zavadsky}, {Pettini}, {Lopez}, {Dall'aglio}, \&
  {Becker}}]{2009ApJ...691..152C}
{Chen}, H.-W., {Perley}, D.~A., {Pollack}, L.~K., {et~al.} 2009, \apj, 691, 152

\bibitem[{{Chornock} {et~al.}(2013){Chornock}, {Berger}, {Fox}, {Lunnan},
  {Drout}, {Fong}, {Laskar}, \& {Roth}}]{2013ApJ...774...26C}
{Chornock}, R., {Berger}, E., {Fox}, D.~B., {et~al.} 2013, \apj, 774, 26

\bibitem[{{Christensen} {et~al.}(2014){Christensen}, {M{\o}ller}, {Fynbo}, \&
  {Zafar}}]{2014MNRAS.445..225C}
{Christensen}, L., {M{\o}ller}, P., {Fynbo}, J.~P.~U., \& {Zafar}, T. 2014,
  \mnras, 445, 225

\bibitem[{{Ciardi} \& {Loeb}(2000)}]{2000ApJ...540..687C}
{Ciardi}, B. \& {Loeb}, A. 2000, \apj, 540, 687

\bibitem[{{Cordero} {et~al.}(2014){Cordero}, {Pilachowski}, {Johnson},
  {McDonald}, {Zijlstra}, \& {Simmerer}}]{2014ApJ...780...94C}
{Cordero}, M.~J., {Pilachowski}, C.~A., {Johnson}, C.~I., {et~al.} 2014, \apj,
  780, 94

\bibitem[{{Cucchiara} {et~al.}(2014){Cucchiara}, {Fumagalli}, {Rafelski},
  {Kocevski}, {Prochaska}, {Cooke}, \& {Becker}}]{2014arXiv1408.3578C}
{Cucchiara}, A., {Fumagalli}, M., {Rafelski}, M., {et~al.} 2014, ArXiv
  e-prints, 1408.3578

\bibitem[{{De Cia} {et~al.}(2012){De Cia}, {Ledoux}, {Fox}, {Vreeswijk},
  {Smette}, {Petitjean}, {Bj{\"o}rnsson}, {Fynbo}, {Hjorth}, \&
  {Jakobsson}}]{2012A&A...545A..64D}
{De Cia}, A., {Ledoux}, C., {Fox}, A.~J., {et~al.} 2012, \aap, 545, A64

\bibitem[{{De Cia} {et~al.}(2013){De Cia}, {Ledoux}, {Savaglio}, {Schady}, \&
  {Vreeswijk}}]{2013A&A...560A..88D}
{De Cia}, A., {Ledoux}, C., {Savaglio}, S., {Schady}, P., \& {Vreeswijk}, P.~M.
  2013, \aap, 560, A88

\bibitem[{{D'Elia} {et~al.}(2014){D'Elia}, {Fynbo}, {Goldoni}, {Covino}, {de
  Ugarte Postigo}, {Ledoux}, \& {Calura}}]{delia2014}
{D'Elia}, V., {Fynbo}, J.~P.~U., {Goldoni}, P., {et~al.} 2014, \aap

\bibitem[{{Denissenkov} \& {Hartwick}(2014)}]{2014MNRAS.437L..21D}
{Denissenkov}, P.~A. \& {Hartwick}, F.~D.~A. 2014, \mnras, 437, L21

\bibitem[{{Dessauges-Zavadsky} {et~al.}(2003){Dessauges-Zavadsky},
  {P{\'e}roux}, {Kim}, {D'Odorico}, \& {McMahon}}]{2003MNRAS.345..447D}
{Dessauges-Zavadsky}, M., {P{\'e}roux}, C., {Kim}, T.-S., {D'Odorico}, S., \&
  {McMahon}, R.~G. 2003, \mnras, 345, 447

\bibitem[{{Deutsch}(1999)}]{1999AJ....118.1882D}
{Deutsch}, E.~W. 1999, \aj, 118, 1882

\bibitem[{{Evans} {et~al.}(2009){Evans}, {Beardmore}, {Page}, {Osborne},
  {O'Brien}, {Willingale}, {Starling}, {Burrows}, {Godet}, {Vetere}, {Racusin},
  {Goad}, {Wiersema}, {Angelini}, {Capalbi}, {Chincarini}, {Gehrels}, {Kennea},
  {Margutti}, {Morris}, {Mountford}, {Pagani}, {Perri}, {Romano}, \&
  {Tanvir}}]{2009MNRAS.397.1177E}
{Evans}, P.~A., {Beardmore}, A.~P., {Page}, K.~L., {et~al.} 2009, \mnras, 397,
  1177

\bibitem[{{Evans} {et~al.}(2007){Evans}, {Beardmore}, {Page}, {Tyler},
  {Osborne}, {Goad}, {O'Brien}, {Vetere}, {Racusin}, {Morris}, {Burrows},
  {Capalbi}, {Perri}, {Gehrels}, \& {Romano}}]{2007A&A...469..379E}
{Evans}, P.~A., {Beardmore}, A.~P., {Page}, K.~L., {et~al.} 2007, \aap, 469,
  379

\bibitem[{{Fan} {et~al.}(2006){Fan}, {Strauss}, {Becker}, {White}, {Gunn},
  {Knapp}, {Richards}, {Schneider}, {Brinkmann}, \&
  {Fukugita}}]{2006AJ....132..117F}
{Fan}, X., {Strauss}, M.~A., {Becker}, R.~H., {et~al.} 2006, \aj, 132, 117

\bibitem[{{Fox} {et~al.}(2008){Fox}, {Ledoux}, {Vreeswijk}, {Smette}, \&
  {Jaunsen}}]{2008A&A...491..189F}
{Fox}, A.~J., {Ledoux}, C., {Vreeswijk}, P.~M., {Smette}, A., \& {Jaunsen},
  A.~O. 2008, \aap, 491, 189

\bibitem[{{Fox} {et~al.}(2007){Fox}, {Petitjean}, {Ledoux}, \&
  {Srianand}}]{2007A&A...465..171F}
{Fox}, A.~J., {Petitjean}, P., {Ledoux}, C., \& {Srianand}, R. 2007, \aap, 465,
  171

\bibitem[{{Fruchter} {et~al.}(2006){Fruchter}, {Levan}, {Strolger},
  {Vreeswijk}, {Thorsett}, {Bersier}, {Burud}, {Castro Cer{\'o}n},
  {Castro-Tirado}, {Conselice}, {Dahlen}, {Ferguson}, {Fynbo}, {Garnavich},
  {Gibbons}, {Gorosabel}, {Gull}, {Hjorth}, {Holland}, {Kouveliotou}, {Levay},
  {Livio}, {Metzger}, {Nugent}, {Petro}, {Pian}, {Rhoads}, {Riess}, {Sahu},
  {Smette}, {Tanvir}, {Wijers}, \& {Woosley}}]{2006Natur.441..463F}
{Fruchter}, A.~S., {Levan}, A.~J., {Strolger}, L., {et~al.} 2006, \nat, 441,
  463

\bibitem[{{Fynbo} {et~al.}(2008){Fynbo}, {Prochaska}, {Sommer-Larsen},
  {Dessauges-Zavadsky}, \& {M{\o}ller}}]{2008ApJ...683..321F}
{Fynbo}, J.~P.~U., {Prochaska}, J.~X., {Sommer-Larsen}, J.,
  {Dessauges-Zavadsky}, M., \& {M{\o}ller}, P. 2008, \apj, 683, 321

\bibitem[{{Gehrels} {et~al.}(2004){Gehrels}, {Chincarini}, {Giommi}, {Mason},
  {Nousek}, {Wells}, {White}, {Barthelmy}, {Burrows}, {Cominsky}, {Hurley},
  {Marshall}, {M{\'e}sz{\'a}ros}, {Roming}, {Angelini}, {Barbier}, {Belloni},
  {Campana}, {Caraveo}, {Chester}, {Citterio}, {Cline}, {Cropper}, {Cummings},
  {Dean}, {Feigelson}, {Fenimore}, {Frail}, {Fruchter}, {Garmire}, {Gendreau},
  {Ghisellini}, {Greiner}, {Hill}, {Hunsberger}, {Krimm}, {Kulkarni}, {Kumar},
  {Lebrun}, {Lloyd-Ronning}, {Markwardt}, {Mattson}, {Mushotzky}, {Norris},
  {Osborne}, {Paczynski}, {Palmer}, {Park}, {Parsons}, {Paul}, {Rees},
  {Reynolds}, {Rhoads}, {Sasseen}, {Schaefer}, {Short}, {Smale}, {Smith},
  {Stella}, {Tagliaferri}, {Takahashi}, {Tashiro}, {Townsley}, {Tueller},
  {Turner}, {Vietri}, {Voges}, {Ward}, {Willingale}, {Zerbi}, \&
  {Zhang}}]{2004ApJ...611.1005G}
{Gehrels}, N., {Chincarini}, G., {Giommi}, P., {et~al.} 2004, \apj, 611, 1005

\bibitem[{{Giavalisco}(2002)}]{2002ARA&A..40..579G}
{Giavalisco}, M. 2002, \araa, 40, 579

\bibitem[{{Goldoni}(2011)}]{2011AN....332..227G}
{Goldoni}, P. 2011, Astronomische Nachrichten, 332, 227

\bibitem[{{Goto} {et~al.}(2011){Goto}, {Utsumi}, {Hattori}, {Miyazaki}, \&
  {Yamauchi}}]{2011MNRAS.415L...1G}
{Goto}, T., {Utsumi}, Y., {Hattori}, T., {Miyazaki}, S., \& {Yamauchi}, C.
  2011, \mnras, 415, L1

\bibitem[{{Gratton} {et~al.}(2001){Gratton}, {Bonifacio}, {Bragaglia},
  {Carretta}, {Castellani}, {Centurion}, {Chieffi}, {Claudi}, {Clementini},
  {D'Antona}, {Desidera}, {Fran{\c c}ois}, {Grundahl}, {Lucatello}, {Molaro},
  {Pasquini}, {Sneden}, {Spite}, \& {Straniero}}]{2001A&A...369...87G}
{Gratton}, R.~G., {Bonifacio}, P., {Bragaglia}, A., {et~al.} 2001, \aap, 369,
  87

\bibitem[{{Greiner} {et~al.}(2009){Greiner}, {Kr{\"u}hler}, {Fynbo}, {Rossi},
  {Schwarz}, {Klose}, {Savaglio}, {Tanvir}, {McBreen}, {Totani}, {Zhang}, {Wu},
  {Watson}, {Barthelmy}, {Beardmore}, {Ferrero}, {Gehrels}, {Kann}, {Kawai},
  {Yolda{\c s}}, {M{\'e}sz{\'a}ros}, {Milvang-Jensen}, {Oates}, {Pierini},
  {Schady}, {Toma}, {Vreeswijk}, {Yolda{\c s}}, {Zhang}, {Afonso}, {Aoki},
  {Burrows}, {Clemens}, {Filgas}, {Haiman}, {Hartmann}, {Hasinger}, {Hjorth},
  {Jehin}, {Levan}, {Liang}, {Malesani}, {Pyo}, {Schulze}, {Szokoly}, {Terada},
  \& {Wiersema}}]{2009ApJ...693.1610G}
{Greiner}, J., {Kr{\"u}hler}, T., {Fynbo}, J.~P.~U., {et~al.} 2009, \apj, 693,
  1610

\bibitem[{{Gunn} \& {Peterson}(1965)}]{1965ApJ...142.1633G}
{Gunn}, J.~E. \& {Peterson}, B.~A. 1965, \apj, 142, 1633

\bibitem[{{Hartoog} {et~al.}(2013){Hartoog}, {Wiersema}, {Vreeswijk}, {Kaper},
  {Tanvir}, {Savaglio}, {Berger}, {Chornock}, {Covino}, {D'Elia}, {Flores},
  {Fynbo}, {Goldoni}, {Gomboc}, {Melandri}, {Pozanenko}, {Schaye}, {Postigo},
  \& {Wijers}}]{hartoog2013}
{Hartoog}, O.~E., {Wiersema}, K., {Vreeswijk}, P.~M., {et~al.} 2013, \mnras,
  430, 2739

\bibitem[{{Im} {et~al.}(2013){Im}, {Sung}, \& {Urata}}]{2013GCN..14800...1I}
{Im}, M., {Sung}, H.-I., \& {Urata}, Y. 2013, GRB Coordinates Network, 14800

\bibitem[{{Jelinek} {et~al.}(2013){Jelinek}, {Gorosabel}, {Castro-Tirado},
  {Mottola}, {Hellmich}, {Fernandez-Munoz}, \&
  {Munoz-Martinez}}]{2013GCN..14782...1J}
{Jelinek}, M., {Gorosabel}, J., {Castro-Tirado}, A.~J., {et~al.} 2013, GRB
  Coordinates Network, 14782, 1

\bibitem[{{Jenkins}(2009)}]{jenkins2009}
{Jenkins}, E.~B. 2009, \apj, 700, 1299

\bibitem[{{Kawai} {et~al.}(2006){Kawai}, {Kosugi}, {Aoki}, {Yamada}, {Totani},
  {Ohta}, {Iye}, {Hattori}, {Aoki}, {Furusawa}, {Hurley}, {Kawabata},
  {Kobayashi}, {Komiyama}, {Mizumoto}, {Nomoto}, {Noumaru}, {Ogasawara},
  {Sato}, {Sekiguchi}, {Shirasaki}, {Suzuki}, {Takata}, {Tamagawa}, {Terada},
  {Watanabe}, {Yatsu}, \& {Yoshida}}]{2006Natur.440..184K}
{Kawai}, N., {Kosugi}, G., {Aoki}, K., {et~al.} 2006, \nat, 440, 184

\bibitem[{{Kraft} {et~al.}(1997){Kraft}, {Sneden}, {Smith}, {Shetrone},
  {Langer}, \& {Pilachowski}}]{1997AJ....113..279K}
{Kraft}, R.~P., {Sneden}, C., {Smith}, G.~H., {et~al.} 1997, \aj, 113, 279

\bibitem[{{Kr{\"u}hler} {et~al.}(2011){Kr{\"u}hler}, {Greiner}, {Schady},
  {Savaglio}, {Afonso}, {Clemens}, {Elliott}, {Filgas}, {Gruber}, {Kann},
  {Klose}, {K{\"u}pc{\"u}-Yolda{\c s}}, {McBreen}, {Olivares}, {Pierini},
  {Rau}, {Rossi}, {Nardini}, {Nicuesa Guelbenzu}, {Sudilovsky}, \&
  {Updike}}]{2011A&A...534A.108K}
{Kr{\"u}hler}, T., {Greiner}, J., {Schady}, P., {et~al.} 2011, \aap, 534, A108

\bibitem[{{Lamb} \& {Reichart}(2000)}]{2000ApJ...536....1L}
{Lamb}, D.~Q. \& {Reichart}, D.~E. 2000, \apj, 536, 1

\bibitem[{{Ledoux} {et~al.}(2006){Ledoux}, {Petitjean}, {Fynbo}, {M{\o}ller},
  \& {Srianand}}]{2006A&A...457...71L}
{Ledoux}, C., {Petitjean}, P., {Fynbo}, J.~P.~U., {M{\o}ller}, P., \&
  {Srianand}, R. 2006, \aap, 457, 71

\bibitem[{{Littlejohns} {et~al.}(2014){Littlejohns}, {Butler}, {Cucchiara},
  {Watson}, {Kutyrev}, {Lee}, {Richer}, {Klein}, {Fox}, {Prochaska}, {Bloom},
  {Troja}, {Ramirez-Ruiz}, {de Diego}, {Georgiev}, {Gonz{\'a}lez},
  {Rom{\'a}n-Z{\'u}{\~n}iga}, {Gehrels}, \& {Moseley}}]{2014AJ....148....2L}
{Littlejohns}, O.~M., {Butler}, N.~R., {Cucchiara}, A., {et~al.} 2014, \aj,
  148, 2

\bibitem[{{Lodders} {et~al.}(2009){Lodders}, {Palme}, \&
  {Gail}}]{2009LanB...4B...44L}
{Lodders}, K., {Palme}, H., \& {Gail}, H.-P. 2009, Landolt B{\"o}rnstein, 44

\bibitem[{{Lunnan} {et~al.}(2013){Lunnan}, {Drout}, {Chornock}, \&
  {Berger}}]{2013GCN..14798...1L}
{Lunnan}, R., {Drout}, M., {Chornock}, R., \& {Berger}, E. 2013, GRB
  Coordinates Network, 14798

\bibitem[{{Meiring} {et~al.}(2009){Meiring}, {Lauroesch}, {Kulkarni},
  {P{\'e}roux}, {Khare}, \& {York}}]{2009MNRAS.397.2037M}
{Meiring}, J.~D., {Lauroesch}, J.~T., {Kulkarni}, V.~P., {et~al.} 2009, \mnras,
  397, 2037

\bibitem[{{Miralda-Escud\'e}(1998)}]{1998ApJ...501...15M}
{Miralda-Escud\'e}, J. 1998, \apj, 501, 15

\bibitem[{{M{\o}ller} {et~al.}(2013){M{\o}ller}, {Fynbo}, {Ledoux}, \&
  {Nilsson}}]{2013MNRAS.430.2680M}
{M{\o}ller}, P., {Fynbo}, J.~P.~U., {Ledoux}, C., \& {Nilsson}, K.~K. 2013,
  \mnras, 430, 2680

\bibitem[{{Monet} {et~al.}(1998){Monet}, A., {Canzian}, {Dahn}, {Guetter},
  {Harris}, {Henden}, {Levine}, {Luginbuhl}, {Monet}, {Rhodes}, {Riepe},
  {Sell}, {Stone}, {Vrba}, \& {Walker}}]{monet1998}
{Monet}, D., A., B., {Canzian}, B., {et~al.} 1998, U.S. Naval Observatory,
  Washington DC

\bibitem[{{Morgan}(2013)}]{2013GCN..14802...1M}
{Morgan}, A.~N. 2013, GRB Coordinates Network, 14802

\bibitem[{{Nagayama}(2013{\natexlab{a}})}]{2013GCN..14794...1N}
{Nagayama}, T. 2013{\natexlab{a}}, GRB Coordinates Network, 14794

\bibitem[{{Nagayama}(2013{\natexlab{b}})}]{2013GCN..14784...1N}
{Nagayama}, T. 2013{\natexlab{b}}, GRB Coordinates Network, 14784

\bibitem[{{Neeleman} {et~al.}(2013){Neeleman}, {Wolfe}, {Prochaska}, \&
  {Rafelski}}]{2013ApJ...769...54N}
{Neeleman}, M., {Wolfe}, A.~M., {Prochaska}, J.~X., \& {Rafelski}, M. 2013,
  \apj, 769, 54

\bibitem[{{Oh} \& {Furlanetto}(2005)}]{2005ApJ...620L...9O}
{Oh}, S.~P. \& {Furlanetto}, S.~R. 2005, \apjl, 620, L9

\bibitem[{{Ono} {et~al.}(2012){Ono}, {Ouchi}, {Mobasher}, {Dickinson},
  {Penner}, {Shimasaku}, {Weiner}, {Kartaltepe}, {Nakajima}, {Nayyeri},
  {Stern}, {Kashikawa}, \& {Spinrad}}]{2012ApJ...744...83O}
{Ono}, Y., {Ouchi}, M., {Mobasher}, B., {et~al.} 2012, \apj, 744, 83

\bibitem[{{Osborne} {et~al.}(2013){Osborne}, {Beardmore}, {Evans}, \&
  {Goad}}]{2013GCN..14811...1O}
{Osborne}, J.~P., {Beardmore}, A.~P., {Evans}, P.~A., \& {Goad}, M.~R. 2013,
  GRB Coordinates Network, 14811

\bibitem[{{Overzier} {et~al.}(2009){Overzier}, {Guo}, {Kauffmann}, {De Lucia},
  {Bouwens}, \& {Lemson}}]{2009MNRAS.394..577O}
{Overzier}, R.~A., {Guo}, Q., {Kauffmann}, G., {et~al.} 2009, \mnras, 394, 577

\bibitem[{{Patel} {et~al.}(2010){Patel}, {Warren}, {Mortlock}, \&
  {Fynbo}}]{2010A&A...512L...3P}
{Patel}, M., {Warren}, S.~J., {Mortlock}, D.~J., \& {Fynbo}, J.~P.~U. 2010,
  \aap, 512, L3

\bibitem[{{Pei}(1992)}]{1992ApJ...395..130P}
{Pei}, Y.~C. 1992, \apj, 395, 130

\bibitem[{{Pentericci} {et~al.}(2014){Pentericci}, {Vanzella}, {Fontana},
  {Castellano}, {Treu}, {Mesinger}, {Dijkstra}, {Grazian}, {Bradac},
  {Cristiani}, {Galametz}, {Giavalisco}, {Giallongo}, {Maiolino}, {Paris}, \&
  {Santini}}]{2014arXiv1403.5466P}
{Pentericci}, L., {Vanzella}, E., {Fontana}, A., {et~al.} 2014, ArXiv e-prints

\bibitem[{{Phillips} {et~al.}(1982){Phillips}, {Gondhalekar}, \&
  {Pettini}}]{phillips1982}
{Phillips}, A.~P., {Gondhalekar}, P.~M., \& {Pettini}, M. 1982, \mnras, 200,
  687

\bibitem[{{Planck Collaboration} {et~al.}(2014){Planck Collaboration}, {Ade},
  {Aghanim}, {Armitage-Caplan}, {Arnaud}, {Ashdown}, {Atrio-Barandela},
  {Aumont}, {Baccigalupi}, {Banday}, \& et~al.}]{2014A&A...571A..16P}
{Planck Collaboration}, {Ade}, P.~A.~R., {Aghanim}, N., {et~al.} 2014, \aap,
  571, A16

\bibitem[{{Portegies Zwart} {et~al.}(2004){Portegies Zwart}, {Baumgardt},
  {Hut}, {Makino}, \& {McMillan}}]{2004Natur.428..724P}
{Portegies Zwart}, S.~F., {Baumgardt}, H., {Hut}, P., {Makino}, J., \&
  {McMillan}, S.~L.~W. 2004, \nat, 428, 724

\bibitem[{{Price} {et~al.}(2007){Price}, {Songaila}, {Cowie}, {Bell Burnell},
  {Berger}, {Cucchiara}, {Fox}, {Hook}, {Kulkarni}, {Penprase}, {Roth}, \&
  {Schmidt}}]{2007ApJ...663L..57P}
{Price}, P.~A., {Songaila}, A., {Cowie}, L.~L., {et~al.} 2007, \apjl, 663, L57

\bibitem[{{Prochaska} {et~al.}(2007{\natexlab{a}}){Prochaska}, {Chen},
  {Dessauges-Zavadsky}, \& {Bloom}}]{2007ApJ...666..267P}
{Prochaska}, J.~X., {Chen}, H.-W., {Dessauges-Zavadsky}, M., \& {Bloom}, J.~S.
  2007{\natexlab{a}}, \apj, 666, 267

\bibitem[{{Prochaska} {et~al.}(2008){Prochaska}, {Dessauges-Zavadsky},
  {Ramirez-Ruiz}, \& {Chen}}]{2008ApJ...685..344P}
{Prochaska}, J.~X., {Dessauges-Zavadsky}, M., {Ramirez-Ruiz}, E., \& {Chen},
  H.-W. 2008, \apj, 685, 344

\bibitem[{{Prochaska} {et~al.}(2002){Prochaska}, {Henry}, {O'Meara}, {Tytler},
  {Wolfe}, {Kirkman}, {Lubin}, \& {Suzuki}}]{2002PASP..114..933P}
{Prochaska}, J.~X., {Henry}, R.~B.~C., {O'Meara}, J.~M., {et~al.} 2002, \pasp,
  114, 933

\bibitem[{{Prochaska} \& {Wolfe}(1997)}]{1997ApJ...487...73P}
{Prochaska}, J.~X. \& {Wolfe}, A.~M. 1997, \apj, 487, 73

\bibitem[{{Prochaska} {et~al.}(2007{\natexlab{b}}){Prochaska}, {Wolfe}, {Howk},
  {Gawiser}, {Burles}, \& {Cooke}}]{2007ApJS..171...29P}
{Prochaska}, J.~X., {Wolfe}, A.~M., {Howk}, J.~C., {et~al.} 2007{\natexlab{b}},
  \apjs, 171, 29

\bibitem[{{Rafelski} {et~al.}(2014){Rafelski}, {Neeleman}, {Fumagalli},
  {Wolfe}, \& {Prochaska}}]{2014ApJ...782L..29R}
{Rafelski}, M., {Neeleman}, M., {Fumagalli}, M., {Wolfe}, A.~M., \&
  {Prochaska}, J.~X. 2014, \apjl, 782, L29

\bibitem[{{Rafelski} {et~al.}(2012){Rafelski}, {Wolfe}, {Prochaska},
  {Neeleman}, \& {Mendez}}]{2012ApJ...755...89R}
{Rafelski}, M., {Wolfe}, A.~M., {Prochaska}, J.~X., {Neeleman}, M., \&
  {Mendez}, A.~J. 2012, \apj, 755, 89

\bibitem[{{Ruiz-Velasco} {et~al.}(2007){Ruiz-Velasco}, {Swan}, {Troja},
  {Malesani}, {Fynbo}, {Starling}, {Xu}, {Aharonian}, {Akerlof}, {Andersen},
  {Ashley}, {Barthelmy}, {Bersier}, {Castro Cer{\'o}n}, {Castro-Tirado},
  {Gehrels}, {G{\"o}{\v g}{\"u}{\c s}}, {Gorosabel}, {Guidorzi}, {G{\"u}ver},
  {Hjorth}, {Horns}, {Huang}, {Jakobsson}, {Jensen}, {K{\i}z{\i}lo{\v g}lu},
  {Kouveliotou}, {Krimm}, {Ledoux}, {Levan}, {Marsh}, {McKay}, {Melandri},
  {Milvang-Jensen}, {Mundell}, {O'Brien}, {{\"O}zel}, {Phillips}, {Quimby},
  {Rowell}, {Rujopakarn}, {Rykoff}, {Schaefer}, {Sollerman}, {Tanvir},
  {Th{\"o}ne}, {Urata}, {Vestrand}, {Vreeswijk}, {Watson}, {Wheeler}, {Wijers},
  {Wren}, {Yost}, {Yuan}, {Zhai}, \& {Zheng}}]{2007ApJ...669....1R}
{Ruiz-Velasco}, A.~E., {Swan}, H., {Troja}, E., {et~al.} 2007, \apj, 669, 1

\bibitem[{{Salvaterra} {et~al.}(2009){Salvaterra}, {Della Valle}, {Campana},
  {Chincarini}, {Covino}, {D'Avanzo}, {Fern{\'a}ndez-Soto}, {Guidorzi},
  {Mannucci}, {Margutti}, {Th{\"o}ne}, {Antonelli}, {Barthelmy}, {de Pasquale},
  {D'Elia}, {Fiore}, {Fugazza}, {Hunt}, {Maiorano}, {Marinoni}, {Marshall},
  {Molinari}, {Nousek}, {Pian}, {Racusin}, {Stella}, {Amati}, {Andreuzzi},
  {Cusumano}, {Fenimore}, {Ferrero}, {Giommi}, {Guetta}, {Holland}, {Hurley},
  {Israel}, {Mao}, {Markwardt}, {Masetti}, {Pagani}, {Palazzi}, {Palmer},
  {Piranomonte}, {Tagliaferri}, \& {Testa}}]{2009Natur.461.1258S}
{Salvaterra}, R., {Della Valle}, M., {Campana}, S., {et~al.} 2009, \nat, 461,
  1258

\bibitem[{{Salvaterra} {et~al.}(2013){Salvaterra}, {Maio}, {Ciardi}, \&
  {Campisi}}]{2013MNRAS.429.2718S}
{Salvaterra}, R., {Maio}, U., {Ciardi}, B., \& {Campisi}, M.~A. 2013, \mnras,
  429, 2718

\bibitem[{{Savage} \& {Sembach}(1996)}]{savage1996}
{Savage}, B.~D. \& {Sembach}, K.~R. 1996, \araa, 34, 279

\bibitem[{{Savaglio}(2001)}]{savaglio2001}
{Savaglio}, S. 2001, in IAU Symposium, Vol. 204, The Extragalactic Infrared
  Background and its Cosmological Implications, ed. M.~{Harwit} \& M.~G.
  {Hauser}, 307

\bibitem[{{Savaglio} {et~al.}(2009){Savaglio}, {Glazebrook}, \& {Le
  Borgne}}]{2009ApJ...691..182S}
{Savaglio}, S., {Glazebrook}, K., \& {Le Borgne}, D. 2009, \apj, 691, 182

\bibitem[{{Schlafly} \& {Finkbeiner}(2011)}]{2011ApJ...737..103S}
{Schlafly}, E.~F. \& {Finkbeiner}, D.~P. 2011, \apj, 737, 103

\bibitem[{{Schlegel} {et~al.}(1998){Schlegel}, {Finkbeiner}, \&
  {Davis}}]{1998ApJ...500..525S}
{Schlegel}, D.~J., {Finkbeiner}, D.~P., \& {Davis}, M. 1998, \apj, 500, 525

\bibitem[{{Songaila}(2004)}]{2004AJ....127.2598S}
{Songaila}, A. 2004, \aj, 127, 2598

\bibitem[{{Sparre} {et~al.}(2014){Sparre}, {Hartoog}, {Kr{\"u}hler}, {Fynbo},
  {Watson}, {Wiersema}, {D'Elia}, {Zafar}, {Afonso}, {Covino}, {de Ugarte
  Postigo}, {Flores}, {Goldoni}, {Greiner}, {Hjorth}, {Jakobsson}, {Kaper},
  {Klose}, {Levan}, {Malesani}, {Milvang-Jensen}, {Nardini}, {Piranomonte},
  {Sollerman}, {S{\'a}nchez-Ram{\'{\i}}rez}, {Schulze}, {Tanvir}, {Vergani}, \&
  {Wijers}}]{2014ApJ...785..150S}
{Sparre}, M., {Hartoog}, O.~E., {Kr{\"u}hler}, T., {et~al.} 2014, \apj, 785,
  150

\bibitem[{{Stark} {et~al.}(2010){Stark}, {Ellis}, {Chiu}, {Ouchi}, \&
  {Bunker}}]{2010MNRAS.408.1628S}
{Stark}, D.~P., {Ellis}, R.~S., {Chiu}, K., {Ouchi}, M., \& {Bunker}, A. 2010,
  \mnras, 408, 1628

\bibitem[{{Steidel} {et~al.}(2003){Steidel}, {Adelberger}, {Shapley},
  {Pettini}, {Dickinson}, \& {Giavalisco}}]{2003ApJ...592..728S}
{Steidel}, C.~C., {Adelberger}, K.~L., {Shapley}, A.~E., {et~al.} 2003, \apj,
  592, 728

\bibitem[{{Tanvir} {et~al.}(2009){Tanvir}, {Fox}, {Levan}, {Berger},
  {Wiersema}, {Fynbo}, {Cucchiara}, {Kr{\"u}hler}, {Gehrels}, {Bloom},
  {Greiner}, {Evans}, {Rol}, {Olivares}, {Hjorth}, {Jakobsson}, {Farihi},
  {Willingale}, {Starling}, {Cenko}, {Perley}, {Maund}, {Duke}, {Wijers},
  {Adamson}, {Allan}, {Bremer}, {Burrows}, {Castro-Tirado}, {Cavanagh}, {de
  Ugarte Postigo}, {Dopita}, {Fatkhullin}, {Fruchter}, {Foley}, {Gorosabel},
  {Kennea}, {Kerr}, {Klose}, {Krimm}, {Komarova}, {Kulkarni}, {Moskvitin},
  {Mundell}, {Naylor}, {Page}, {Penprase}, {Perri}, {Podsiadlowski}, {Roth},
  {Rutledge}, {Sakamoto}, {Schady}, {Schmidt}, {Soderberg}, {Sollerman},
  {Stephens}, {Stratta}, {Ukwatta}, {Watson}, {Westra}, {Wold}, \&
  {Wolf}}]{2009Natur.461.1254T}
{Tanvir}, N.~R., {Fox}, D.~B., {Levan}, A.~J., {et~al.} 2009, \nat, 461, 1254

\bibitem[{{Tanvir} {et~al.}(2012){Tanvir}, {Levan}, {Fruchter}, {Fynbo},
  {Hjorth}, {Wiersema}, {Bremer}, {Rhoads}, {Jakobsson}, {O'Brien}, {Stanway},
  {Bersier}, {Natarajan}, {Greiner}, {Watson}, {Castro-Tirado}, {Wijers},
  {Starling}, {Misra}, {Graham}, \& {Kouveliotou}}]{2012ApJ...754...46T}
{Tanvir}, N.~R., {Levan}, A.~J., {Fruchter}, A.~S., {et~al.} 2012, \apj, 754,
  46

\bibitem[{{Th{\"o}ne} {et~al.}(2013){Th{\"o}ne}, {Fynbo}, {Goldoni}, {de
  Ugarte}, {Campana}, {Vergani}, {Covino}, {Kr{\"u}hler}, {Kaper}, {Tanvir},
  {Zafar}, {D'Elia}, {Gorosabel}, {Greiner}, {Groot}, {Hammer}, {Jakobsson},
  {Klose}, {Levan}, {Milvang-Jensen}, {Nicuesa}, {Palazzi}, {Piranomonte},
  {Tagliaferri}, {Watson}, {Wiersema}, \& {Wijers}}]{2013MNRAS.428.3590T}
{Th{\"o}ne}, C.~C., {Fynbo}, J.~P.~U., {Goldoni}, P., {et~al.} 2013, \mnras,
  428, 3590

\bibitem[{{Totani} {et~al.}(2014){Totani}, {Aoki}, {Hattori}, {Kosugi},
  {Niino}, {Hashimoto}, {Kawai}, {Ohta}, {Sakamoto}, \&
  {Yamada}}]{2014PASJ...66...63T}
{Totani}, T., {Aoki}, K., {Hattori}, T., {et~al.} 2014, \pasj, 66, 63

\bibitem[{{Totani} {et~al.}(2006){Totani}, {Kawai}, {Kosugi}, {Aoki}, {Yamada},
  {Iye}, {Ohta}, \& {Hattori}}]{2006PASJ...58..485T}
{Totani}, T., {Kawai}, N., {Kosugi}, G., {et~al.} 2006, \pasj, 58, 485

\bibitem[{{Ukwatta} {et~al.}(2013){Ukwatta}, {Barthelmy}, {Gehrels}, {Krimm},
  {Malesani}, {Marshall}, {Maselli}, {Melandri}, {Palmer}, \&
  {Stamatikos}}]{2013GCN..14781...1U}
{Ukwatta}, T.~N., {Barthelmy}, S.~D., {Gehrels}, N., {et~al.} 2013, GRB
  Coordinates Network, 14781

\bibitem[{{Utsumi} {et~al.}(2010){Utsumi}, {Goto}, {Kashikawa}, {Miyazaki},
  {Komiyama}, {Furusawa}, \& {Overzier}}]{2010ApJ...721.1680U}
{Utsumi}, Y., {Goto}, T., {Kashikawa}, N., {et~al.} 2010, \apj, 721, 1680

\bibitem[{{Vacca} {et~al.}(2003){Vacca}, {Cushing}, \& {Rayner}}]{vacca03}
{Vacca}, W.~D., {Cushing}, M.~C., \& {Rayner}, J.~T. 2003, \pasp, 115, 389

\bibitem[{{Vernet} {et~al.}(2011){Vernet}, {Dekker}, {D'Odorico}, {Kaper},
  {Kjaergaard}, {Hammer}, {Randich}, {Zerbi}, {Groot}, {Hjorth}, {Guinouard},
  {Navarro}, {Adolfse}, {Albers}, {Amans}, {Andersen}, {Andersen}, {Binetruy},
  {Bristow}, {Castillo}, {Chemla}, {Christensen}, {Conconi}, {Conzelmann},
  {Dam}, {de Caprio}, {de Ugarte Postigo}, {Delabre}, {di Marcantonio},
  {Downing}, {Elswijk}, {Finger}, {Fischer}, {Flores}, {Fran{\c c}ois},
  {Goldoni}, {Guglielmi}, {Haigron}, {Hanenburg}, {Hendriks}, {Horrobin},
  {Horville}, {Jessen}, {Kerber}, {Kern}, {Kiekebusch}, {Kleszcz}, {Klougart},
  {Kragt}, {Larsen}, {Lizon}, {Lucuix}, {Mainieri}, {Manuputy}, {Martayan},
  {Mason}, {Mazzoleni}, {Michaelsen}, {Modigliani}, {Moehler}, {M{\o}ller},
  {Norup S{\o}rensen}, {N{\o}rregaard}, {P{\'e}roux}, {Patat}, {Pena}, {Pragt},
  {Reinero}, {Rigal}, {Riva}, {Roelfsema}, {Royer}, {Sacco}, {Santin},
  {Schoenmaker}, {Spano}, {Sweers}, {Ter Horst}, {Tintori}, {Tromp}, {van
  Dael}, {van der Vliet}, {Venema}, {Vidali}, {Vinther}, {Vola}, {Winters},
  {Wistisen}, {Wulterkens}, \& {Zacchei}}]{2011A&A...536A.105V}
{Vernet}, J., {Dekker}, H., {D'Odorico}, S., {et~al.} 2011, \aap, 536, A105

\bibitem[{{Vladilo} {et~al.}(2001){Vladilo}, {Centuri{\'o}n}, {Bonifacio}, \&
  {Howk}}]{2001ApJ...557.1007V}
{Vladilo}, G., {Centuri{\'o}n}, M., {Bonifacio}, P., \& {Howk}, J.~C. 2001,
  \apj, 557, 1007

\bibitem[{{Vreeswijk} {et~al.}(2004){Vreeswijk}, {Ellison}, {Ledoux}, {Wijers},
  {Fynbo}, {M{\o}ller}, {Henden}, {Hjorth}, {Masi}, {Rol}, {Jensen}, {Tanvir},
  {Levan}, {Castro Cer{\'o}n}, {Gorosabel}, {Castro-Tirado}, {Fruchter},
  {Kouveliotou}, {Burud}, {Rhoads}, {Masetti}, {Palazzi}, {Pian}, {Pedersen},
  {Kaper}, {Gilmore}, {Kilmartin}, {Buckle}, {Seigar}, {Hartmann}, {Lindsay},
  \& {van den Heuvel}}]{2004A&A...419..927V}
{Vreeswijk}, P.~M., {Ellison}, S.~L., {Ledoux}, C., {et~al.} 2004, \aap, 419,
  927

\bibitem[{{Vreeswijk} {et~al.}(2013){Vreeswijk}, {Ledoux}, {Raassen}, {Smette},
  {De Cia}, {Wo{\'z}niak}, {Fox}, {Vestrand}, \&
  {Jakobsson}}]{2013A&A...549A..22V}
{Vreeswijk}, P.~M., {Ledoux}, C., {Raassen}, A.~J.~J., {et~al.} 2013, \aap,
  549, A22

\bibitem[{{Vreeswijk} {et~al.}(2007){Vreeswijk}, {Ledoux}, {Smette}, {Ellison},
  {Jaunsen}, {Andersen}, {Fruchter}, {Fynbo}, {Hjorth}, {Kaufer}, {M{\o}ller},
  {Petitjean}, {Savaglio}, \& {Wijers}}]{vreeswijk2007}
{Vreeswijk}, P.~M., {Ledoux}, C., {Smette}, A., {et~al.} 2007, \aap, 468, 83

\bibitem[{{Wang} {et~al.}(2012){Wang}, {Bromm}, {Greif}, {Stacy}, {Dai},
  {Loeb}, \& {Cheng}}]{2012ApJ...760...27W}
{Wang}, F.~Y., {Bromm}, V., {Greif}, T.~H., {et~al.} 2012, \apj, 760, 27

\bibitem[{{Wang} \& {Dai}(2014)}]{2014ApJS..213...15W}
{Wang}, F.~Y. \& {Dai}, Z.~G. 2014, \apjs, 213, 15

\bibitem[{{Watson}(2011)}]{Watson11}
{Watson}, D. 2011, \aap, 533, A16

\bibitem[{{Wijers} {et~al.}(1998){Wijers}, {Bloom}, {Bagla}, \&
  {Natarajan}}]{1998MNRAS.294L..13W}
{Wijers}, R.~A.~M.~J., {Bloom}, J.~S., {Bagla}, J.~S., \& {Natarajan}, P. 1998,
  \mnras, 294, L13

\bibitem[{{Wolfe} {et~al.}(2005){Wolfe}, {Gawiser}, \&
  {Prochaska}}]{2005ARA&A..43..861W}
{Wolfe}, A.~M., {Gawiser}, E., \& {Prochaska}, J.~X. 2005, \araa, 43, 861

\bibitem[{{Wolfe} {et~al.}(2003){Wolfe}, {Prochaska}, \&
  {Gawiser}}]{2003ApJ...593..215W}
{Wolfe}, A.~M., {Prochaska}, J.~X., \& {Gawiser}, E. 2003, \apj, 593, 215

\bibitem[{{Xu} {et~al.}(2013{\natexlab{a}}){Xu}, {Malesani}, {Schulze},
  {Fynbo}, {D'Elia}, {Goldoni}, {Hartoog}, {Hjorth}, {Kaper}, {Kruehler},
  {Levan}, {Milvang-Jensen}, {Tanvir}, \& {Wiersema}}]{2013GCN..14816...1X}
{Xu}, D., {Malesani}, D., {Schulze}, S., {et~al.} 2013{\natexlab{a}}, GRB
  Coordinates Network, 14816

\bibitem[{{Xu} {et~al.}(2013{\natexlab{b}}){Xu}, {Malesani}, {Schulze},
  {Tanvir}, {Watson}, {Hjorth}, {Datson}, \& {Salinas}}]{2013GCN..14783...1X}
{Xu}, D., {Malesani}, D., {Schulze}, S., {et~al.} 2013{\natexlab{b}}, GRB
  Coordinates Network, 14783

\bibitem[{{Xu} \& {Wei}(2009)}]{2009ScChG..52.1428X}
{Xu}, Z. \& {Wei}, D. 2009, Science in China G: Physics and Astronomy, 52, 1428

\bibitem[{{Zafar} {et~al.}(2011){Zafar}, {Watson}, {Fynbo}, {Malesani},
  {Jakobsson}, \& {de Ugarte Postigo}}]{2011A&A...532A.143Z}
{Zafar}, T., {Watson}, D., {Fynbo}, J.~P.~U., {et~al.} 2011, \aap, 532, A143

\end{thebibliography}
\appendix
\section{Full absorption spectra}
\begin{figure*}
\centering
\includegraphics[width=\textwidth]{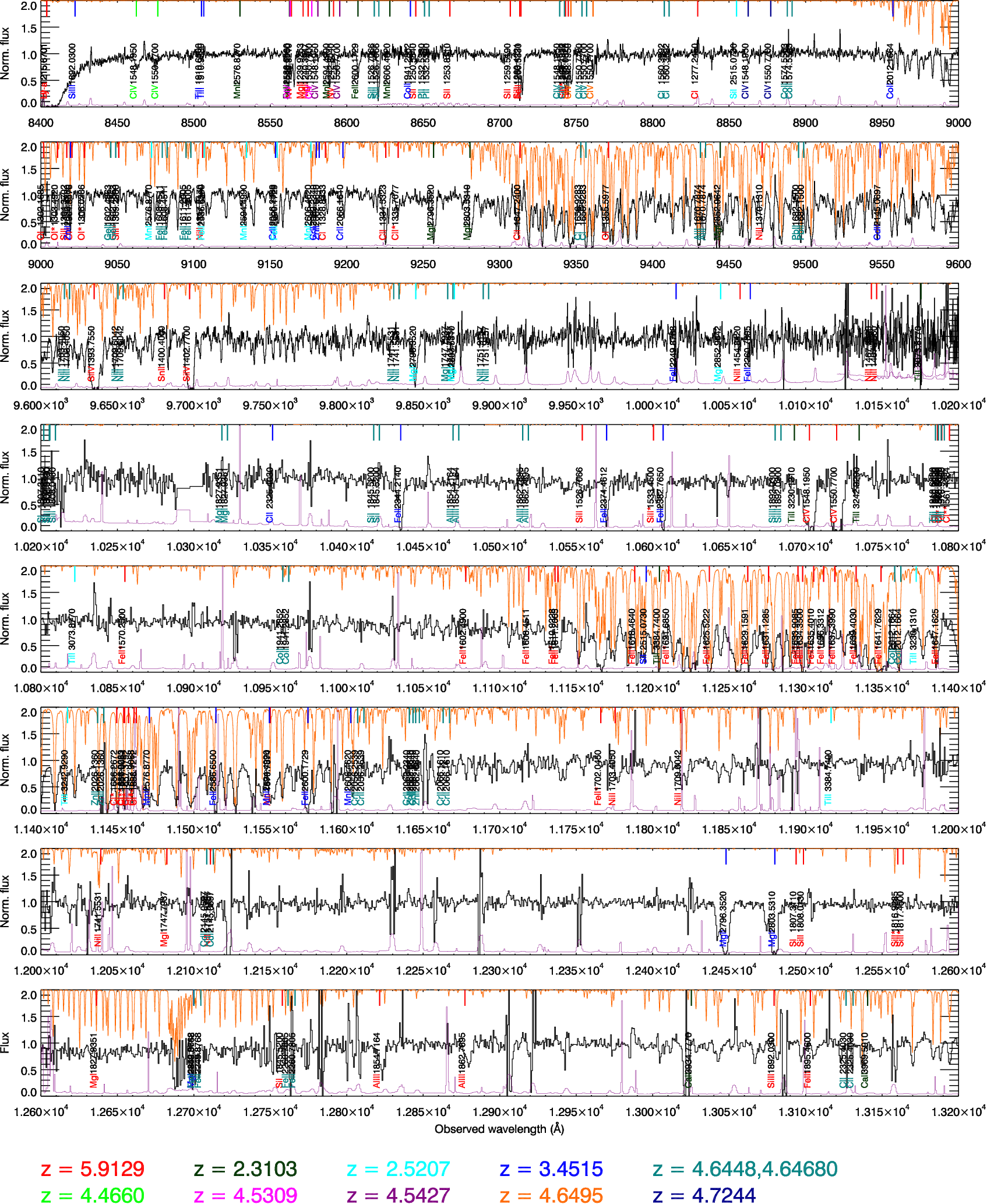}
\caption{Afterglow spectrum op \thisgrb, starting from Ly$\alpha$. The absorption lines are indicated with the ion that produces them and the rest wavelength. We use different colours for the various absorbers (see Sect.~\ref{sec:int}), as indicated in the legend; $z=5.9127$ in red is the signature of the host galaxy. The error spectrum is shown in purple, the orange spectrum is the scaled atmospheric transmission spectrum. The spectrum continues in Fig.~\ref{fig:spec_part2}. }
\label{fig:spec_part1}
\end{figure*}

\begin{figure*}
\centering
\includegraphics[width=\textwidth]{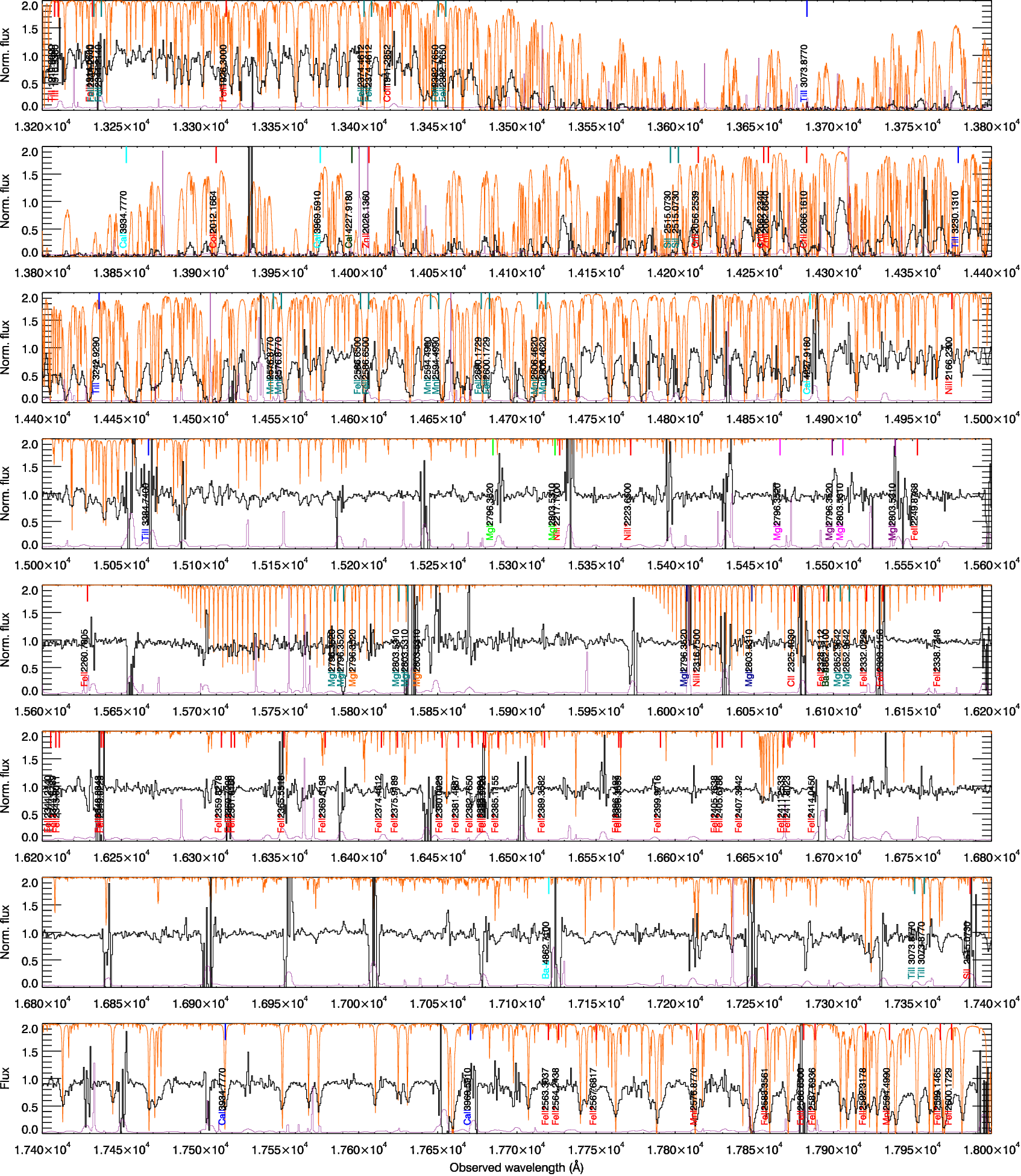}
\caption{Continuation of Fig.~\ref{fig:spec_part1}}
\label{fig:spec_part2}
\end{figure*}

\end{document}